\documentclass[%
preprintnumbers,
nofootinbib,
 amsmath,amssymb,
 aps,
 a4paper,
prd,
]{revtex4-2}

\usepackage{caption}
\usepackage[utf8]{inputenc}
\usepackage{fullpage}
\usepackage{bbm,braket}
\usepackage{slashed, float,tabu, booktabs,mathrsfs}
\usepackage{tikz,graphicx,subcaption}
\usepackage[compat=1.1.0]{tikz-feynman}
\usepackage[hidelinks]{hyperref}

\usepackage{subcaption}
\usepackage{physics}

\usepackage{xcolor}

\def\bea{\begin{eqnarray}}
\def\eea{\end{eqnarray}}

\newcommand{\mL}{\mathcal{L}}

\newcommand{\mM}{\mathcal{M}}
\newcommand{\mO}{\mathcal{O}}

\newcommand{\vecbf}[1]{\Vec{\mathbf{#1}}}
\newcommand{\F}{\mathring{F}_\pi}

\newcommand{\no}{\nonumber}

\newlength{\myfigwidth}
\begin{document}

\title{Hyperon non-leptonic decays\\in relativistic Chiral Perturbation Theory with resonances}
\author{Nora Salone$^1$}
\email{nora.salone@us.edu.pl}
\author{Fernando Alvarado$^{2}$}
\email{f.alvarado@gsi.de}
\author{Stefan Leupold$^3$}
\email{stefan.leupold@physics.uu.se}
\author{Andrzej Kupsc$^{3,4}$}
\email{andrzej.kupsc@physics.uu.se}

\affiliation{$^1$ University of Silesia in Katowice, Institute of Physics, \\ 75 Pulku Piechoty 1, 41-500 Chorzow, Poland}
\affiliation{$^2$ GSI Helmholtzzentrum f\"ur Schwerionenforschung GmbH, \\ Planckstraße 1, 64291 
Darmstadt, Germany}
\affiliation{$^3$ Institutionen f\"or fysik och astronomi, Uppsala universitet, Box 516, S-75120 Uppsala, Sweden}
\affiliation{$^4$ National Centre for Nuclear Research,  Pasteura 7, 02-093 Warsaw, Poland}
\date{\today}

\begin{abstract} 
Motivated by recent experimental advances in the corresponding measurements, non-leptonic hyperon decays are calculated, for the first time in a relativistic manner, in Chiral Perturbation Theory at next-to-leading order (NLO). On the one hand, relativistic loop corrections are computed explicitly based on the ground-state octet and decuplet fields. On the other hand, the NLO weak-transition low-energy constants are estimated by resonance saturation, inspired by the non-relativistic tree-level computation of Ref.~\cite{Borasoy:1999md}. In particular, the $1/2^-$ and the (excited) $1/2^+$ resonance octets are utilized. The remaining unknown parameters are fitted to the decay amplitudes. A good combined fit to both $s$- and $p$-wave amplitudes is achieved with the caveat of not being very tightly constrained. The role of the resonances is found to be crucial. Consequences for further investigations and open questions are addressed.
\end{abstract}

\maketitle

\section{Introduction}
\label{sec:intro}

The non-leptonic weak decays of hyperons consist of the following seven processes: 
$\Sigma^+  \rightarrow n \, \pi^+$,  
$\Sigma^+  \rightarrow p \, \pi^0$,  
$\Sigma^- \rightarrow n \, \pi^-$,  
$\Lambda \rightarrow p \, \pi^-$,
$\Lambda \rightarrow n \, \pi^0$,
$\Xi^- \rightarrow  \Lambda \, \pi^-$, and 
$\Xi^0 \rightarrow  \Lambda \, \pi^0 $. They decompose in an {\it s}- and {\it p}-wave amplitude of odd and even parity, respectively.\footnote{We follow the tradition from atomic spectroscopy and call the partial wave of orbital angular momentum $L=0$ the {\it s}-wave (sharp). The $L=1$ wave is denoted by {\it p} (principal).} The description of these processes constitutes a long standing theoretical problem\footnote{It is worth commenting also on the numerical approach to the question at hand, that is to say on the prospect of lattice QCD computations. Although two-body meson-baryon states are currently studied on the lattice in scattering computations, a number of significant challenges arise in the computation of weak non-leptonic decays. Relevant results in this direction have been obtained in the meson sector for $K\to\pi\pi$, notice for instance Ref.~\cite{RBC:2023ynh}.}; the present study builds upon many works that have analysed these decays in Chiral Perturbation Theory (ChPT) at different levels of sophistication and precision, and under slightly different assumptions. In the past, all works had in common that the non-relativistic heavy-baryon framework has been used.

At leading order (LO) in ChPT one cannot describe both {\it s}- and {\it p}-waves simultaneously \cite{SM-Donoghue}, and at next-to-leading order (NLO) the description of the two waves has been problematic ever since the first study of Ref.~\cite{Bijnens:1985kj}. In that work only the non-analytic terms ($M_K^2\log \left(M_K^2/\mu^2\right)$) of the chiral corrections were considered, arguing that the chiral logs could in principle be larger than the analytic $M_K^2$ terms. The results exhibit a poor agreement with data, with the {\it p}-wave contributions being particularly large. Afterwards, the study of Ref.~\cite{Jenkins:1991bt} performed an analogous estimation with the addition of decuplet baryons as explicit degrees of freedom. This made it possible to describe the {\it s}-waves with a satisfying convergence pattern, but not the {\it p}-waves. The same qualitative findings have also been obtained more recently in a framework \cite{Flores-Mendieta:2019lao} that supplements chiral perturbation theory by a $1/N_c$ expansion, where $N_c$ denotes the number of colors.

Hand in hand with the convergence problems of the chiral series comes the fact that too many unknown weak-transition low-energy constants (LECs) contribute to the analytic $\mO(M_K^2)$ term. There are too few decay amplitudes to fix all of them. 
Previous works on the topic include a variety of different approaches: in Ref.~\cite{Borasoy:1998ku}, the LECs (counterterms) contributing to $\mO(M_K^2)$ are included to a certain extent, while the decuplet is integrated out. In Ref.~\cite{Borasoy:1999md} the calculation is performed at tree level and the impact of baryon resonances on the counterterms is assessed. In particular, the $1/2^-$ multiplet, containing $N(1535)$, and the $1/2^+$ Roper-like octet were included. The authors conclude that the effect of these resonances is beneficial to the chiral convergence of the series, however they call for a higher-order calculation including loops in order to make a ``more quantitative statement''. One purpose of the present work is to fill this gap.

Recently the BESIII collaboration measured non-leptonic hyperon decays yielding results for the polarization asymmetry~\cite{BESIII:2018cnd,BESIII:2021ypr}, i.e. the {\it s}- and {\it p}-wave interference term, considerably different from previous values (see~\cite{Salone:2022lpt} and references therein for an overview of the experimental situation). 
In light of these experimental updates, we feel motivated to attempt also a theoretical update. 

For ChPT applications in the sector of the strong interaction, relativistic calculations have become state of the art. Therefore we compute the hyperon non-leptonic decays for the first time in a relativistic manner at NLO $\mathcal{O}(M_K^2)$ in ChPT. The decuplet baryons are introduced as explicit degrees of freedom, as they are expected to contribute significantly via loop diagrams. Since the inclusion of relativistic baryons in ChPT leads to the well-known breaking of the power counting of the theory, an appropriate renormalization scheme must be chosen. Here the power-counting-breaking terms are subtracted following the extended-on-mass-shell renormalization scheme (EOMS)~\cite{Fuchs:2003qc, Geng:2013xn}, preserving power counting, as well as covariance and the analytic properties of the theory. 

Of course, this does not solve the problem that at NLO several unknown weak transition LECs contribute. Inspired by Refs.\ \cite{Borasoy:1999md,Borasoy:1999nt}, we approximate all NLO counterterms by including the tree-level contributions from the $1/2^-$ and the (Roper) $1/2^+$ resonant baryon octets. In other words, LECs are generated by integrating out these resonances. Potentially existing other LECs are neglected. This follows the spirit of resonance saturation introduced in \cite{Ecker:1988te,Donoghue:1988ed} with large success in the strong-interaction sector \cite{Ecker:1989yg,Epelbaum:2001fm,Kampf:2011ty,Bijnens:2014lea}. After computing the amplitudes, we fit the unknown weak LECs to the updated experimental data and discuss the quality of the description. 

One aspect of the updated input concerns the decomposition of the measured decay amplitudes into $\Delta I=1/2$ and $\Delta I=3/2$ parts. We deduce explicitly the $\Delta I=1/2$ parts utilizing also recent pion-nucleon final-state phase shifts \cite{Hoferichter:2015hva}. As much as possible, we address only the $\Delta I=1/2$ parts by our ChPT calculation. Thus we will \emph{not} fit to the complete experimental amplitudes that are to some extent polluted by $\Delta I=3/2$ contributions. We regard such a more detailed analysis as timely in view of the data quality and the already mentioned revision of old values for the decay asymmetries. After fitting the relativistic ChPT expressions with their NLO accuracy of $\mathcal{O}(M_K^2)$ to the decomposed decay amplitudes, we discuss the consequences of our results for further developments.

The paper is structured as follows: first, the theoretical framework pertaining our description of hyperon non-leptonic decays is presented (Sec.\ \ref{sec:rel-BChPT}). Afterwards, the careful extraction of the decay amplitudes from experimental data to be compared to theory is explained (Sec.~\ref{sec:expt-ampl-fit}), followed by the technical details of the ChPT calculation (Sec.~\ref{sec:perturbative calculation}). Then the results from the fit to data are discussed (Sec.~\ref{sec:fit}). The main text ends with a conclusion and outlook (Sec.\ \ref{sec:Conclusion-and-outlook}). Several appendices are added to address technical details that would interrupt the main text. 

\section{Hyperon non-leptonic decays in relativistic ChPT}
\subsection{Relativistic Baryon ChPT}\label{sec:rel-BChPT}
We set out to compute the decay amplitudes at $\mO(M_K^2)$ in ChPT. To do so, the following effective Lagrangian is needed:
\begin{equation}\label{eq:Leff}
    \mL_{\rm eff}=\mL_{\phi}+\mL_{\phi BT}+\mL_{\phi RB}
    +\mL^{\rm W}_{\phi BT}+\mL^{W}_{RB}  \ ,
\end{equation}
where the first three terms on the right-hand side correspond to standard strong-interaction ChPT and the last two parameterize the weak transition. The field content is indicated by $\phi$, $B$ and $T$, referring to the lowest-lying mesons (Goldstone bosons), the baryon octet and the baryon decuplet, respectively. The index $R$ represents the resonance fields in general: the $1/2^-$ octet and the excited $1/2^+$ octet. Decomposing in chiral orders, the above terms read:
\begin{equation}\label{eq:Lagrangians}
    \mL_{\phi}= \mL^{(2)}_{\phi}+\mL^{(4)}_{\phi}\ , \quad \mL_{\phi B T}= \mL^{(1)}_{\phi B T} + \ldots \ , \quad \mL_{\phi RB}= \mL^{(1)}_{\phi RB} + \ldots \ , \quad\ \mL^W_{\phi B T}= \mL^{W(0)}_{\phi B T} \ , \quad \mL^W_{RB}= \mL^{W(0)}_{RB} \ .
\end{equation}
The superscript numbers denote the respective order in the chiral power counting. An ellipsis denotes higher-order terms where an explicit specification is not needed for the present work. This refers, in particular, to the mass splittings in flavor multiplets, which appear at second order in the chiral counting (first order in the quark masses), i.e.\ at the same order as the loop diagrams. Baryon mass differences are included in all diagrams, introducing higher-order contributions while respecting the power counting. The power counting is discussed further in Sec.~\ref{sec:perturbative calculation}. Higher-order terms of the weak baryon Lagrangian are not reported since their counterterms are approximated by the contribution of the $1/2^{\pm}$ resonances, as will be discussed in Sec. \ref{sec:resonances}.

\subsubsection{Lowest-lying hadrons\label{sec:lowest lying hadrons}}
The mesonic Lagrangians $\mL^{(2)}_{\phi}$, $\mL^{(4)}_{\phi}$ are standard and given in Ref.~\cite{Gasser:1984gg}. We briefly describe the lowest-order relativistic chiral Lagrangian for the ground-state octet and decuplet baryons \cite{Jenkins:1991bt,Jenkins:1991ts,Holmberg:2018dtv,Lutz:2001yb,Semke:2005sn,Pascalutsa:2006up,Geng:2008mf,Mommers:2022dgw}:
    \begin{eqnarray}
        \mL^{(1)}_{\phi BT} &=& \tr\left[ \bar{B} \left( i\slashed{\mathcal{D}} - m_B \right) B \right] + \frac{D}{2} \tr\left( \bar{B} \gamma^\mu  \gamma_5 \left\{ u_\mu, B \right\} \right) + \frac{F}{2} \tr\left( \bar{B} \gamma^\mu \gamma_5 \left[ u_\mu, B \right] \right)
\nonumber \\*
        &&{}        + \bar{T}^\mu_{abc} \left[ i \gamma_{\mu\nu\alpha} \left(\mathcal{D}^\alpha T^\nu \right)^{abc} - \gamma_{\mu\nu} m_{T} \left( T^\nu \right)^{abc} \right] + \frac{\mathcal{H}}{2} \bar{T}^\mu_{abc} \gamma_\nu \gamma_5 \left(u^\nu\right)^c_d T^{abd}_\mu   \nonumber \\*
        &&{} + \frac{\mathcal{C}}{2} \left( \epsilon^{ade} \bar{T}^\mu_{abc} \left(u_\mu\right)^b_d B^c_e + \epsilon_{ade} \bar{B}^e_c \left( u^\mu \right)^d_b T^{abc}_\mu \right) \,, 
\label{eq:stronglag}
    \end{eqnarray}
where tr denotes a flavor trace. We have introduced totally antisymmetric products of gamma matrices,
\begin{eqnarray}
    \gamma^{\mu\nu} &=& \frac{1}{2}\left[ \gamma^\mu, \gamma^\nu \right] = -i\sigma^{\mu\nu}, \\*
    \gamma^{\mu\nu\alpha} &=& \frac{1}{2} \left\{ \gamma^{\mu\nu}, \gamma^\alpha \right\} \,. 
\end{eqnarray}

The fields of the ground-state octet baryons are collected in a $3 \times 3$ flavor matrix with matrix elements $B^a_b$. Contrary to Ref.~\cite{Jenkins:1991bt}, we use {\it a} to count the rows and {\it b} for the columns of the matrix (cf.\ also the discussion in \cite{Mommers:2022dgw}). The matrix is given by
\begin{equation}
  B = (B^a_b) = \left(
    \begin{array}{ccc}
      \frac{1}{\sqrt{2}} \Sigma^0 + \frac{1}{\sqrt{6}} \Lambda & \Sigma^+ & p \\
      \Sigma^- & -\frac{1}{\sqrt{2}} \Sigma^0 + \frac{1}{\sqrt{6}} \Lambda & n  \\
      \Xi^- & \Xi^0 & -\frac{2}{\sqrt{6}} \Lambda
    \end{array}
  \right)   \,.
    \label{eq:3Bgs}
\end{equation}
The baryon decuplet is introduced in the so-called ``small scale expansion'' power counting~\cite{Hemmert:1997ye}, which assigns the same counting to octet and decuplet propagators. The decuplet baryons are expressed via a completely symmetric third-rank tensor $T^{abc}$ with 
\begin{eqnarray}
  &T^{111} = \Delta^{++} \,, \quad &T^{112} = \frac{1}{\sqrt{3}} \Delta^+ \,, \nonumber \\* 
  &T^{122} = \frac{1}{\sqrt{3}} \Delta^0 \,, \quad &T^{222} = \Delta^-  \,, \nonumber \\*
  &T^{113} = \frac{1}{\sqrt{3}} \Sigma^{\ast +} \,, \quad &T^{123} = \frac{1}{\sqrt{6}} \Sigma^{\ast 0} \,,
    \quad T^{223} = \frac{1}{\sqrt{3}} \Sigma^{\ast -} \,,  \nonumber \\*
  &T^{133} = \frac{1}{\sqrt{3}} \Xi^{\ast 0} \,, \quad &T^{233} = \frac{1}{\sqrt{3}} \Xi^{\ast -} \,, \quad  T^{333} = \Omega^- \,.  
  \label{eq:decuplet-states}
\end{eqnarray}
As for Lorentz-covariant structures, the decuplet baryons are spin-3/2 vector-spinors, while the octet baryons are spin-1/2 spinors, whose indices are not spelled out in Eq.~\eqref{eq:3Bgs} and Eq.~\eqref{eq:decuplet-states}. 

The meson fields are encoded in the vector and axial-vector combinations
\begin{equation}
    \Gamma^\mu = \frac{1}{2}\left( u \partial^\mu u^\dagger + u^\dagger \partial^\mu u \right), \quad u^\mu = i\left( u \partial^\mu u^\dagger - u^\dagger \partial^\mu u \right) \ ,
\end{equation}
with
\begin{equation}
    u = \exp{\frac{i\phi}{\F}}, 
\end{equation}
and $\phi$ the meson octet matrix
\begin{equation}
    \phi = \frac{1}{\sqrt{2}}\left( \begin{array}{ccc}
      \frac{1}{\sqrt{2}}\pi^0 + \frac{1}{\sqrt{6}}\eta  & \pi^+ & K^+ \\
      \pi^-  & -\frac{1}{\sqrt{2}}\pi^0 + \frac{1}{\sqrt{6}}\eta & K^0 \\
       K^-  & \Bar{K}^0 &-\frac{2}{\sqrt{6}}\eta
    \end{array} \right) \ .
    \label{eq:meson-field-matrix}
\end{equation}
The covariant derivatives for the octet and decuplet kinetic term are defined as~\cite{Mommers:2022dgw}
\begin{equation}
    \begin{split}
        &\mathcal{D}^\mu B = \partial^\mu B + [\Gamma^\mu, B] \\ 
        &(\mathcal{D}^\mu T_\nu)^{abc} = \partial^\mu T_\nu^{abc} + (\Gamma^\mu)^a_d T_\nu^{dbc} + (\Gamma^\mu)^b_d T_\nu^{adc} + (\Gamma^\mu)^c_d T_\nu^{abd} \ .
    \end{split} 
\end{equation}

Turning now to the weak transition, the four-quark effective Lagrangian relevant for the $s\to d$ ($\Delta S=1$) process reads
\begin{equation}
    \mL^{\Delta S = 1} = \frac{4G_F}{\sqrt{2}} V_{ud} V_{us}^* \Bar{u}\gamma^\mu P_L s \ \Bar{d} \gamma_\mu P_L u \ ,
\label{eq:four-quark-weak}
\end{equation}
with $G_F$ the Fermi constant, $V$ the CKM matrix,  $P_L$ the left-handed chiral projection operator.
The corresponding weak transition Lagrangian for baryons reads~\cite{Jenkins:1991bt}:
\begin{equation}\label{eq:weaklag}
    \begin{split}
        \mL^{W(0)}_{\phi B T} = h_D \tr \left(\Bar{B}\{ h_+ , B\} \right) + h_F \tr \left(\Bar{B}[ h_+ , B]\right) + h_C \Bar{T}^\mu_{abc} \, (h_+)^c_d \,  T_\mu^{abd}  \ ,
    \end{split} 
\end{equation}
with $h_+=u^\dagger (h + h^\dagger) u$ and
\begin{equation}
    h = \left( \begin{array}{ccc}
        0 & 0 & 0 \\
        0 & 0 & 1 \\ 
        0 & 0 & 0
    \end{array} \right) \ .
\end{equation}

The weak Lagrangian (\ref{eq:weaklag}) allows for two-point vertices where a hyperon fluctuates directly into another baryon with a different strangeness content. These transitions are parity conserving, contributing finally to {\it p}-wave amplitudes. Chiral symmetry breaking relates such a matrix element to the process with an additional soft pion. The corresponding three-point vertices, which emerge from (\ref{eq:weaklag}), contribute to parity violating {\it s}-wave amplitudes. Naively one would think that at low energies, the {\it p}-wave amplitude is suppressed relative to the {\it s}-wave by a momentum factor. However, the direct fluctuation of one baryon to another gives rise to a propagator where the mass difference between the two baryons appears in the denominator (see also Figs.\ \ref{1b}, \ref{1c}, \ref{3a}-\ref{3t} below). This mass difference is essentially proportional to the strange-quark mass $m_s$. The corresponding enhancement by $1/m_s \sim 1/M_K^2$ promotes the {\it p}-wave such that it becomes comparable to the {\it s}-wave \cite{Jenkins:1991bt}.

Note that (\ref{eq:weaklag}) is not the most general weak chiral Lagrangian of leading (zeroth) order. Following common practice \cite{Jenkins:1991bt,Cirigliano:2011ny}, we assume that the weak transition operator (encoded via $h$) transforms as an octet and not as a 27-plet with respect to left-handed chiral transformations. In practice, this implies that the strangeness changing transition is accompanied by an isospin change of $\Delta I=1/2$. In contrast, a transition caused by a 27-plet operator contains also a $\Delta I=3/2$ part. Four-quark operators cannot lead to larger isospin changes. The subleading 27-plet transitions have been studied within heavy-baryon ChPT in \cite{AbdEl-Hady:1998qww,Tandean:1998ch}. We focus here on the octet transitions. Detailed analyses in the kaon sector suggest that in the standard model the $\Delta I=1/2$ transitions are numerically dominant by a factor of about 1/20 \cite{Jenkins:1991bt,Cirigliano:2011ny}. Also calculations in lattice QCD support this finding \cite{RBC:2020kdj,RBC:2023ynh}. We assume that the same holds in the hyperon sector and restrict the effective Lagrangian to the octet-transition terms given in (\ref{eq:weaklag}).

Also the phenomenology of the non-leptonic hyperon decays fits well with the assumption that the $\Delta I=1/2$ transitions are numerically dominant, though the $\Delta I=3/2$ contribution is not negligible and particularly big, albeit subleading, for the decays of the $\Omega$ baryon \cite{ParticleDataGroup:2024cfk,BESIII:2023jhj}. The decay amplitudes are decomposed in Sec.\ \ref{sec:expt-ampl-fit} and Appendix \ref{app:isospin-etc} into their $\Delta I=1/2$ and $\Delta I=3/2$ parts. In principle, such a decomposition leads to two solutions: one where the $\Delta I=1/2$ amplitude is dominant and another solution where the $\Delta I=3/2$ amplitude is the largest. For instance, let's have a brief look at the decay widths of $\Lambda \to p \pi^-$ and $\Lambda \to n \pi^0$. Their ratio is given by $\approx 1.8$ \cite{ParticleDataGroup:2024cfk}. A pure $\Delta I=1/2$ amplitude would yield a ratio of exactly 2. The natural conclusion is that the $\Delta I=1/2$ amplitude is dominant. But formally, there are two solutions. Allowing for a $\Delta I=1/2$ and a $\Delta I=3/2$ amplitude, a fit to the decay widths of $\Lambda \to p \pi^-$ and $\Lambda \to n \pi^0$ allows for the following two solutions for the ratio of the 3/2 to the 1/2 amplitude: $-0.027$ and $+3.1$. In this back-to-the-envelope calculation, we have neglected the (small) final-state interactions, because the purpose of this exercise is just to highlight what phenomenology alone tells us and where additional theory arguments enter. In line with the findings from the kaon sector \cite{Cirigliano:2011ny,RBC:2020kdj,RBC:2023ynh} we prefer the first solution ($-0.027$) and do not attempt to analyze data with an assumption of $\Delta I= 3/2$ dominance.  
  
In previous studies~\cite{Jenkins:1991bt, AbdEl-Hady:1999llb} a three-meson weak transition was also included. Its size can be fixed from kaon decays and therefore does not induce additional unknown parameters. Such a term is of leading (second) order in the purely mesonic sector \cite{Cirigliano:2011ny}. On the other hand, when coupled to the baryon sector, this corresponds to diagrams of higher order than what we consider here, $\mO(M_K^4)$ \cite{Jenkins:1991bt}. We have also checked the numerical impact of the three-meson vertex on our analysis, which turned out to be negligible. Therefore we leave it out of the final analysis.

\subsubsection{\texorpdfstring{$J^P=1/2^{\pm}$}{} resonances\label{sec:resonances}}

We account for the presence of the $1/2^-$ and the (excited) $1/2^+$ octets in line with Refs.~\cite{Borasoy:1999md,Borasoy:1999nt}. The general term ``resonances'' is employed throughout this work only to refer to the aforementioned excited octets. To be precise, on the one hand the $1/2^-$ octet contains the $N(1535)$ and $\Lambda(1670)$, together with  $\Sigma(1680)$ and $\Xi(1810)$ states, which are less well-established~\cite{Wang:2024jyk,Nishibuchi:2022zfo,Crede:2024hur}: we denote states belonging to this multiplet as $R_{-}$. On the other hand, the $1/2^+$ octet of excited states includes the Roper state $N(1440)$ together with $\Lambda(1600)$, $\Sigma(1660)$ and $\Xi(1790)$, which have been shown to be relevant for low-energy dynamics (see for instance Ref.~\cite{Banerjee:1995wz,Beane:2002ud}). These fields of positive parity, denoted as $R_+$, contribute to the parity conserving {\it p}-wave amplitudes of the non-leptonic decays. Conversely, the $1/2^-$ octet fields $R_{-}$ contribute to the {\it s}-wave. For both multiplets the existence and mass of the cascade states $\Xi$ are not fully established \cite{ParticleDataGroup:2024cfk}. Here the mass of the respective $\Xi$-like state is estimated using the Gell-Mann--Okubo mass relation \cite{Gell-Mann:1961omu, Okubo:1961jc, Okubo:1962zzc}
\begin{equation} \label{eq:gell-okubo}
    m_\Xi \simeq \frac{3 m_\Lambda-2m_N + m_\Sigma}{2} \, .
\end{equation}

The resonance degrees of freedom are implemented in the Lagrangians in Eq.~\eqref{eq:Lagrangians}, $\mL^{(1)}_{\phi RB}=\mL^{(1)}_{\phi R_{-}B}+\mL^{(1)}_{\phi R_+B}$ and $\mL^{W(0)}_{\phi RB}=\mL^{W(0)}_{\phi R_{-}B}+\mL^{W(0)}_{\phi R_+B}$. From Ref.~\cite{Borasoy:1999md}, the Lagrangians responsible for the strong resonant interaction, respecting the symmetries of ChPT, are
\begin{equation}\label{eq:L1phiRB}
    \mL^{(1)}_{\phi R_{-}B}=i \tr \left(\Bar{R}_{-}\slashed{\mathcal{D}}R_{-}\right) - m_{R_{-}} \tr \left(\Bar{R}_{-}R_{-} \right)+ \left[i s_d \ {\rm tr} \left( \bar{R}_{-}\gamma_\mu \{u^\mu,B\} \right) + is_f \ {\rm tr} \left(\bar{R}_{-}\gamma_\mu [u^\mu,B]\right) + {\rm h.c.}\right] \, , 
\end{equation}
and 
\begin{eqnarray}
    \mL^{(1)}_{\phi R_+ B}&=& i \tr \left(\Bar{R}_+\slashed{\mathcal{D}}R_+\right) - m_{R_+} \tr \left(\Bar{R}_+ R_+\right) \nonumber \\
    && {}+ \left[ \frac{D^*}{4} \tr \left(\Bar{R}_+\gamma^\mu\gamma_5 \{u_\mu,B\}\right) + \frac{F^*}{4} \tr \left(\Bar{R}_+\gamma^\mu\gamma_5 [u_\mu,B]\right)  +{\rm h.c.} \right] \, . 
    \label{eq:L1phiB*B}
\end{eqnarray}
Also from Ref.~\cite{Borasoy:1999md}, the corresponding weak transition Lagrangians are
\begin{equation}\label{eq:weak-s-res-L}
    \mL^{W(0)}_{R_{-}B}=
    iw_d \ {\rm tr}\left(\bar R_{-} \{h_+,B\}  \right) 
    +iw_f \ {\rm tr}\left(\bar R_{-} [h_+,B]  \right)
+ {\rm h.c.}
\, , 
\end{equation}
and
\begin{equation}\label{eq:weak-p-res-L}
    \mL^{W(0)}_{R_+B}= d^* \ {\rm tr}\left(\bar R_+ \{h_+,B\}  \right) 
    +f^* \ {\rm tr}\left(\bar R_+ [h_+,B]  \right)
+ {\rm h.c.}
\end{equation}

Also here the weak Lagrangian gives rise to processes where a baryon fluctuates directly into another baryon, in particular a ground-state hyperon into a resonance. For the transition amplitude of the non-leptonic decay (see also Fig.~\ref{1d} below) this leads to a propagator that picks up the difference of the corresponding masses in the denominator, as discussed in detail in Appendix~\ref{sec:res-sat}. We have already discussed a similar process in the paragraph after Eq.\ (\ref{eq:weaklag}). There we identified a $1/m_s$ enhancement factor. From the point of view of the original chiral power counting \cite{Jenkins:1991bt,Scherer:2012xha}, the mass difference between a ground-state baryon (the decaying hyperon) and a resonance is formally not a small scale. This is formally in contrast to the previous situation with the $1/m_s$ chiral enhancement. But numerically, the resonant octets are close in mass to the spin-1/2 ground-state octet, if one compares baryons with {\em different} strangeness. The numerically small values of these mass differences suggest that the contributions from the resonances could be large. For instance, the fluctuation of the ground-state $\Sigma$ baryon to the nucleon induces a mass difference of about $250\,$MeV. This is numerically similar to the fluctuations of the $\Sigma$ to a nucleon resonance: $m_{N(1535)}-m_\Sigma\simeq 346\,$MeV for the $1/2^-$ octet, $m_{N^*(1440)}-m_\Sigma\simeq 250\,$MeV for the $1/2^+$ multiplet. In the present work we do {\em not} modify the chiral power counting. Only mass differences that vanish in the chiral limit are counted as small. But we regard it as reasonable that the resonances contribute significantly to the transition amplitudes. In light of these findings, we include the leading contribution of the excited octets. Our approximation is that such contributions sufficiently account for the NLO counterterms of the Lagrangian of the lowest-lying baryons\footnote{See Ref.~\cite{Borasoy:1998ku} if interested in the NLO counterterms, with the caveat of the theory being non-relativistic (heavy-baryon) in that reference.}. These terms correspond to the diagrams in Figs.~\ref{1d} and~\ref{1e}. In the context of ChPT, the effect of such states has been shown to be relevant in Ref.~\cite{Borasoy:1999md}. However, the aforementioned work differs from the present one in various aspects: it is non-relativistic and also incorporates the resonances only as $1/m_{R}$ contributions, neglecting the ground-state mass in $1/(m_{R}-m_B)$. In view of the numerical smallness of the mass difference---as compared to a baryon mass itself---we follow Ref.\ \cite{Borasoy:1999nt} and keep the mass difference in the denominators.

We stress that we restrict ourselves here to the two lowest-lying resonance octets. Singlets are not considered. In this context we regard the negative-parity $\Lambda$ resonance state(s) in the 1400 MeV mass region \cite{ParticleDataGroup:2024cfk} as a singlet in the quark model, probably with a large or even dominant admixture of meson-baryon components \cite{Siegel:1988rq,Garcia-Recio:2003ejq,Kolomeitsev:2003kt,Magas:2005vu,Mai:2012dt,Hall:2014uca,BaryonScatteringBaSc:2023zvt}. 
For the non-leptonic decays, treated at tree level based on the Lagrangians (\ref{eq:L1phiRB}), (\ref{eq:L1phiB*B}), (\ref{eq:weak-s-res-L}), (\ref{eq:weak-p-res-L}), a $\Lambda$ resonance appears only in the decays of the $\Sigma$ baryons. Here a fluctuation of this $\Lambda$ resonance into a neutron contributes (see Fig.~\ref{1e} below). The corresponding mass difference is not as small as the ones discussed previously, demoting the importance of $\Lambda$ resonances in general and singlet $\Lambda$'s in particular.

\subsection{Hyperon non-leptonic decays}\label{sec:expt-ampl-fit}

The amplitude describing hyperon non-leptonic decays is defined as
\begin{equation}
    i\mathcal{M}(B_i \to B_f \pi) = G_F m_{\pi^+}^2 \overline{u}_{B_f} \{{S} + {P} \gamma_5\} u_{B_i}
    \label{eq:nlamp}
\end{equation}
where the contributions to the transition amplitude are related to the dimensionless partial waves {\it s} and {\it p} via
\begin{equation}
    s = S\quad ; \quad p = P\frac{|\vecbf{p}_f|}{E_f+m_f}  \ . \label{eq:sp-def}
\end{equation}
The {\it s} and {\it p} amplitudes are the object of our theoretical calculations, and can be related to experimental data via the following relations to the decay parameters:
\begin{align}
    \Gamma &= G_F^2 m_{\pi^+}^4 \frac{|\vecbf{p}_f|}{4\pi m_i} \left( E_f + m_f \right) \left(|s|^2 + |p|^2\right) \ , \label{eq:Gamma} \\
    \alpha &= \frac{2\Re\left(s^*p\right)}{|s|^2 + |p|^2} \ , \label{eq:alpha}\\
    \phi   &={\arg}\!\left\{(s+p)(s^* - p^*)\right\}  \ , \label{eq:phi}
\end{align}
with $m_f, \ |\vecbf{p}_f| , \ E_f$ the final-state baryon mass, momentum and energy, respectively, in the rest frame of the mother baryon of mass $m_i$. 

Seven hyperon non-leptonic decays are measurable. Each decay populates two partial waves. All 14 amplitudes can be accessed by data on decay widths (\ref{eq:Gamma}) and decay asymmetries (\ref{eq:alpha},\ref{eq:phi}). In our calculations we restrict ourselves to the parts of the decay amplitudes that adhere to the $\Delta I=1/2$ rule. In this limit, only four decay amplitudes are independent, as isospin symmetry provides the following relations \cite{Jenkins:1991bt}:
\begin{equation}
    \begin{split}
        &\sqrt{2}\mathcal{M}(\Sigma^+ \to p \pi^0) - \mathcal{M}(\Sigma^+ \to n \pi^+) + \mathcal{M}(\Sigma^- \to n \pi^-) = 0 \ , \\ 
        &\mathcal{M}(\Lambda \to p \pi^-) + \sqrt{2}\mathcal{M}(\Lambda \to n \pi^0) = 0 \ , \\ 
        &\mathcal{M}(\Xi^- \to \Lambda \pi^-) + \sqrt{2}\mathcal{M}(\Xi^0 \to \Lambda \pi^0) = 0 \ .
    \end{split}
    \label{eq:deltaI12rule-decays}
\end{equation}

In many previous studies of non-leptonic hyperon decays, data have been utilized as if they satisfied the $\Delta I = 1/2$ relations (\ref{eq:deltaI12rule-decays}). In view of the high quality of recent data, we pursue here a different approach by extracting from the data the respective $\Delta I=1/2$ contributions. 
This approach is particularly meaningful for our calculational scheme. 
Including the resonant contributions via the weak Lagrangians (\ref{eq:weak-s-res-L}), (\ref{eq:weak-p-res-L}) together with the weak Lagrangian for the ground states (\ref{eq:weaklag}), the total number of weak LECs to be estimated by a fit is also seven. It would be misleading to view the procedure as a fit of seven parameters to 14 data points. 
In order to preserve the predictive power of our fit without assuming {\it a priori} Eq.~(\ref{eq:deltaI12rule-decays}), we extract the $\Delta I=1/2$ parts by comparing the pertinent isospin decompositions for each decay. This can be written as
\begin{equation}\label{eq:amp-decomp}
    L_{\rm decay} = \sum_{\Delta I, I} c_{2\Delta I, 2I} L_{2\Delta I, 2I} \exp(i\delta^L_{2I}), \quad \text{with}\ L=S,P§ \text{ and}\ \Delta I, I \in \{1/2,3/2\} \, ,
\end{equation}
where the Clebsch-Gordan coefficients $c_{2\Delta I, 2I}$ can be read off from Eqs.~(\ref{eq:isospin-decomp-Lambda}-\ref{eq:Sisospin2-full}), and the final-state interaction phases are taken from Ref.~\cite{Salone:2022lpt}, following \cite{Hoferichter:2015hva}. In particular, we re-arrange our theoretical amplitudes to perform a direct comparison to eight independent $L_{1,2I}$ `experimental' isospin amplitudes in the $\Delta I = 1/2$ limit.
    \begin{table}[ht]
        \centering
       \[ \begin{array}{llcc}
              &  \Gamma [G_F^2m_{\pi^+}^4 {\rm GeV}]& \langle\alpha\rangle & \phi[^\circ] \\ \toprule
        \Sigma^+\to n\pi^+ & 0.0793 \pm 0.0005 & {0.0506 \pm 0.0032} & 167 \pm 20\\
        \Sigma^+\to p\pi^0 & 0.0792 \pm 0.0004 & -0.9869 \pm 0.0019 & 36 \pm 34\\
        \Sigma^-\to n\pi^- & 0.0861 \pm 0.0006 & -0.068 \pm 0.008 & 10 \pm 15\\
        \Lambda\to p\pi^- & 0.03123 \pm 0.00027 & 0.7542 \pm 0.0026 & -6.5 \pm 3.5\\
        \Lambda\to n\pi^0 & 0.01749 \pm 0.00025 & 0.672 \pm 0.005 & -  \\
        \Xi^-\to \Lambda\pi^- & 0.0777 \pm 0.0008 & {-0.384} \pm 0.002 & -0.1 \pm 0.3\\
        \Xi^0\to \Lambda\pi^0 & 0.0438 \pm 0.0014 & -0.3770 \pm 0.0028 & 0.3 \pm 0.4\\
        \end{array} \]
        \caption{Experimental values of $\Gamma$, $\alpha$ and $\phi$. We include the newest results from BESIII. 
        The remaining results are consistent with the PDG~\cite{ParticleDataGroup:2024cfk}. See the discussion in the main text for more details.
         \label{tab:Input-expt}
     }             
    \end{table}

The $L_{1,2I}$ are estimated in the following way. The recent experimental results for  $\Gamma$, $\alpha$, and $\phi$ are listed in Table~\ref{tab:Input-expt}. The partial decay widths $\Gamma$ are extracted using as input lifetimes and branching fractions as reported by the Particle Data Group (PDG)~\cite{ParticleDataGroup:2024cfk} with the exception of the branching fractions for $\Sigma^+\to p\pi^0$ and  $\Sigma^+\to n\pi^+$ taken from the latest BESIII measurements~\cite{BESIII:2025rgd}. They are given by $49.8(2)\%$ and $49.9(3)\%$, respectively, and exhibit a tension of 3 to 4$\sigma$ with the previous world averages. The $\alpha$ values are extracted from the BESIII measurements: $\Lambda\to p\pi^-$~\cite{BESIII:2022qax}, $\Lambda\to n\pi^0$~\cite{BESIII:2025wxe}, $\Xi^-\to\Lambda\pi^-$ \cite{BESIII:2026hgj},  $\Xi^0\to\Lambda\pi^0$ \cite{BESIII:2023drj}, $\Sigma^+\to p\pi^0$ \cite{BESIII:2025jxt}  and $\Sigma^+\to n\pi^+$~\cite{BESIII:2023sgt} where we use the $\langle\alpha\rangle=(\alpha-\bar\alpha)/2$ hyperon--antihyperon averages instead of $\alpha$ as the former are often much more precise. The $\phi$ values for $\Sigma$ and $\Lambda$ are from the PDG. For $\Xi$ the $\langle\phi\rangle=(\phi-\bar\phi)/2$ averages are taken from BESIII~\cite{BESIII:2026hgj,BESIII:2023drj}. 

We used the data to extract the isospin amplitudes for $\Lambda$, $\Sigma$ and $\Xi$ using three separate sets of fits. In each set, the $\chi^2$ fits were repeated twice to assess the significance of the $\Delta I=3/2$ components. Neglecting these terms results in an over-constrained fit. Therefore, the $\chi^2$ value allows us to test the hypothesis that only the $\Delta I=1/2$ amplitudes are sufficient to describe the data. If the hypothesis is falsified and the $\Delta I=3/2$ contribution is necessary, the central values for the $\Delta I=1/2$ amplitudes are biased. We therefore perform a second fit that includes the $\Delta I=3/2$ components. This fit is unconstrained, meaning that the  $\chi^2$ value is zero but allows one to obtain an unbiased value for the $\Delta I=1/2$ amplitudes relevant for the present study, together with the correct uncertainties obtained from the $\chi^2=1$ contour. Any ambiguities in the solution for the second fit are resolved by requiring that the corrections to the $\Delta I=1/2$-only hypothesis are small. In the following, we discuss the three sets of fits one by one. 
\begin{itemize}
    \item The fit for the $\Lambda$ decays includes the known $N\pi$ strong final-state phase shifts \cite{Salone:2022lpt,Hoferichter:2015hva} but doesn't include $\phi$ since it is not known for $\Lambda\to n\pi^0$. The fit without $\Delta I=3/2$ amplitudes gives $\chi^2/ndf=29.6/2$.\footnote{As it is standard, $ndf$ means number of degrees of freedom.}
    \item The fit for the $\Sigma$ decays includes again the known $N\pi$ strong phases but also the values of the $\phi$ parameters. The fit without $\Delta I=3/2$ amplitudes gives $\chi^2/ndf=105/5$.
    \item  The fit for the $\Xi$ decays doesn't include the $\Lambda\pi$ strong phases since they are not known well \cite{Lu:1994ex,Tandean:2000dx,Meissner:2000re}. 
    An alternative to determine the phase shifts would be to use the decay parameters $\phi$ in the way outlined in \eqref{eq:phiXi}. 
    There one sees that $\phi$ is proportional to the strong phase shift difference $\delta_1^P-\delta_1^S$. The proportionality factors are determined as $-0.42(3)$  and  $-0.41(5)$ for the  $\Xi^-$ and $\Xi^0$ decays, respectively. 
    Unfortunately, the values of $\phi$ are small and the present data situation is such that  they are compatible with zero within their uncertainties. The fit without $\Delta I=3/2$ amplitudes gives $\chi^2/ndf=37.8/2$.
\end{itemize}
The values and errors in Tab.~\ref{tab:SP-expt1}  are for the fits including $\Delta I=3/2$ amplitudes. In the case of $\Lambda$ and $\Xi$ they are not constrained ($ndf=0$) but allow to extract the correct uncertainties. The fit leads to very small $\Delta I=3/2$ amplitudes, which makes it reasonable to use a theory framework with focus on the dominant $\Delta I=1/2$ amplitudes. 
\begin{table}[ht]
        \centering
        \[ \begin{array}{crr}
              &  S \phantom{mmm} & P \phantom{mm} \\ \toprule
             L_{1,1}^\Lambda & 1.412 \pm 0.008 & 11.42 \pm 0.13 \\[0.3em]
             L_{3,3}^\Lambda & -0.045 \pm 0.009 & 0.71 \pm 0.13 \\[0.3em]
             L_{1,1}^\Xi & 2.044 \pm 0.014 & -6.58 \pm 0.10 \\[0.3em]
             L_{3,1}^\Xi & -0.072 \pm 0.018 & 0.27 \pm 0.13 \\[0.3em] 
             L_{1,1}^\Sigma & -0.936 \pm 0.007 & 27.13 \pm 0.09 \\[0.3em]
             L_{1,3}^\Sigma & 1.948 \pm 0.006 & 0.15 \pm 0.07 \\[0.3em]
             L_{3,3}^\Sigma & -0.035 \pm 0.005 & -0.83 \pm 0.06 \\
        \end{array} \]
        \caption{The values for $S$ and $P$ as introduced in (\ref{eq:nlamp}), decomposed in (\ref{eq:amp-decomp}), and finally determined by using the experimental input presented in Tab.~\ref{tab:Input-expt} and the final-state interaction phase shifts $\delta$ for $\pi N$ scattering. Note that $L^\Sigma_{3,1}$ has been neglected; see the corresponding discussion in Appendix \ref{app:isospin-etc}. Also note that the factor $G_F m_{\pi^+}^2$ appears explicitly in (\ref{eq:nlamp}). 
        \label{tab:SP-expt1}}
    \end{table}

The $L_{1,2I}$ amplitudes of Table~\ref{tab:SP-expt1} are the data to which we fit the ChPT amplitudes in our analysis. This constitutes a different approach from previous studies. For instance, in Ref.~\cite{Jenkins:1991bt} the ChPT parametrisation is fitted to the `experimental' amplitudes in the physical basis, not the isospin basis, and phase shifts have been neglected. For the sake of comparison, the `experimental' amplitudes extracted in Ref.~\cite{Jenkins:1991bt} are displayed in Table~\ref{tab:SP-expt} as $l_{\rm old}$ ($l=s,p$). From our results of Tab.~\ref{tab:SP-expt1} and Eq.~\eqref{eq:amp-decomp}, values of the `experimental' amplitudes in the physical basis are computed and displayed in Tab.~\ref{tab:SP-expt} as $\Re(l_{\rm iso})$. Lastly, the real part of the complex-valued amplitudes obtained by solving together Eqs.~(\ref{eq:Gamma}, \ref{eq:alpha}) in the $\Delta I = 1/2$ limit are displayed as $\Re(l_{\rm complex})$. Comparing the different values reveals in some channels significant deviations. This gives further credit to our strategy of using the experimental values of highest quality, those final-state phase shifts that are well known, and a decomposition into the $\Delta I = 1/2$ and $3/2$ parts wherever possible.
    \begin{table}[ht]
        \centering
        \[\begin{array}{lrrrrrr}
              & \Re(s_{\rm iso}) &  \Re(s_{\rm complex}) & s_{\rm old} \phantom{m} & \Re(p_{\rm iso}) & \Re(p_{\rm complex})  & p_{\rm old} \phantom{m}  \\ \toprule
           \Sigma^+\to n\pi^+   & 0.033     & 0.062(07)    & 0.06(01)      & 1.827     & 1.796(09)        & 1.81(01) \\
           \Sigma^+\to p\pi^0   & -1.318  & -1.368(06)     & -1.43(05)   & 1.294     & 1.245(07)        & 1.17(07) \\
           \Sigma^-\to n\pi^-   & 1.897     & 1.848(07)    & 1.88(01)      & -0.003  & -0.064(08)     & -0.06(01) \\
           \Lambda\to p\pi^-    & 1.403     & 1.363(07)    & 1.42(01)      & 0.610     &  0.634(05)      & 0.52(02) \\
           \Lambda\to n\pi^0    & -0.992  & -1.023(10)     & -1.04(01)   & -0.431  & -0.419(13)     & -0.39(04) \\
           \Xi^-\to \Lambda\pi^- & -2.046 & -1.994(09)     & -1.98(01)   & 0.397     & 0.393(05)        & 0.48(02) \\
           \Xi^0\to \Lambda\pi^0 & 1.447    & 1.523(24)    & 1.52(02)      & -0.281  & -0.271(08)    & -0.33(02)
        \end{array} \]
        \caption{Different extractions of the $s$- and $p$-wave `experimental' amplitudes. Note that none of these values constitute the data for our fits. They are displayed only for comparison. $\Re(l_{\rm iso})$ are the amplitudes obtained from the isospin amplitudes in Table \ref{tab:SP-expt1}. On the other hand, $\Re(l_{\rm complex})$ and $l_{\rm old}$ \cite{Jenkins:1991bt} are solutions to the system of equations composed by Eqs.~(\ref{eq:Gamma}, \ref{eq:alpha}) assuming the $\Delta I = 1/2$ limit. See main text for details.}
        \label{tab:SP-expt}
    \end{table}

\subsection{Perturbative calculation\label{sec:perturbative calculation}}
In this section, the perturbative calculation of the {\it s-} and {\it p}-wave amplitudes is discussed. It is an $\mO(M_K^2)$ relativistic ChPT calculation with explicit decuplet fields and including tree-level resonant contributions. The relevant Feynman diagrams are displayed in Figs.~\ref{fig:tree-diags}-\ref{fig:p-wave-diags}. Note that we do not show explicitly those loop diagrams that renormalize the propagators by mass and wave-function renormalization. Those diagrams can be found in the literature, e.g.\ the diagrams of Fig.\ 4 in \cite{Jenkins:1991bt} and diagrams (p) and (q) of Fig.\ 3 in \cite{Borasoy:1998ku}. For the tree-level diagrams we use physical masses, making mass renormalization obsolete. We take care of the wave-function renormalization by appropriate $Z$-factors for all the lines appearing in Figs.\ \ref{1a}, \ref{1b}, \ref{1c}.

The loops are dimensionally regularized at the relevant scale, $\mu=1$ GeV. No counterterms are explicitly included, given that this would introduce a large number of unknown parameters. Instead, the UV divergences are subtracted in the $\widetilde{\rm{MS}}$ scheme \cite{Gasser:1983yg,Gasser:1984gg} while the $\mO(M_K^2)$ counterterms are approximated by the contributions of the resonances at tree-level as explained in App.~\ref{sec:res-sat}. An additional subtraction is required to cancel the power-counting breaking terms caused by the presence of baryon masses in the loops, which is done following the extended-on-mass-shell (EOMS) renormalization prescription~\cite{Fuchs:2003qc}. In this way, power counting is restored without altering the analytic properties of the loops and preserving covariance. In all terms, the differences between baryon masses are kept, even if they constitute corrections beyond the order of the calculation, as their impact can be relevant\footnote{As a note to avoid possible mistakes, we emphasize that one has to correctly identify the order of each mass-splitting contribution and subtract the power breaking terms consequently. To put it explicitly, given a radiative correction with a specific baryon, $B_{i}$, inside the loop, $\mM(m_i)$, one considers the expansion of the baryon mass around the chiral limit quantity, $m_B$, i.e. $m_i=m_{B}+\lambda_{i}M_{\phi}^2$. Then, after the subtraction of the power counting breaking term, the $m_B$ term contributes at leading one-loop order (Eq.~\eqref{eq:amplitude}) and $\lambda_{i}$ only enters two orders higher. See \cite{Ando:2006xy} for a similar discussion.}.  Note that, as customary~\cite{Yao:2016vbz,Alvarado:2022bok}, power-counting breaking terms are identified and subtracted in an expansion in powers of all the light quark masses but not in the decuplet-octet baryon mass difference. 

As a result, we obtain expressions for the amplitudes which depend on a handful of LECs. For every decay, a different amplitude with the following structure is obtained (summation over $\phi=K,\pi,\eta$ and wave-function-renormalization indices $\Phi=\pi,B,B'$ is implied):
\begin{eqnarray}\label{eq:amplitude}
    \mM &=& \frac{1}{\F}\Bigg\{\alpha + f_{\rm res}(M_\phi^2)   \no\\
    &&\left.+\frac{M_\phi^2}{16\pi^2\F^2}\left[\beta^\phi +\gamma^\phi  \log\left(\frac{M_\phi^2}{\mu^2}\right)\right] +\alpha \frac{M_\phi^2}{16\pi^2\F^2}\left[\beta_{Z\Phi}^\phi +\gamma_{Z\Phi}^\phi  \log\left(\frac{M_\phi^2}{\mu^2}\right)\right]+\mO(M_\phi^3)\right\} \ .
\end{eqnarray}
The first term in the curly bracket on the right-hand side is the tree-level plain ChPT contribution, the second term is the resonance contribution, the third term denotes the loops, and the last is the contribution from wave-function renormalization. The quantities $\alpha$, $\beta$, $\gamma$, and $f_{\rm res}$ depend implicitly on the LECs and the baryon masses $m_B$, $m_T$, and the latter also on $m_R$. $\F \simeq 0.0803$ GeV is the value of the chiral-limit pion decay constant including chiral corrections at next order, based on the calculation of Ref.~\cite{Borasoy:1998ku}. The amplitudes were calculated with the help of FeynCalc~\cite{Shtabovenko:2016sxi, Shtabovenko:2020gxv, Mertig:1990an, Shtabovenko:2016whf, Patel:2015tea, Patel:2016fam}, and the amplitude expressions are provided without truncation in a Mathematica notebook as supplementary material.

\subsubsection{Resonance parameters from strong decays}
\label{sec:resonances-strong-decays}

The contributions of the $1/2^-$ and (excited) $1/2^+$ octets to the hyperon non-leptonic decays (Figs.~\ref{1d}, \ref{1e}) introduce the coupling constants related to the strong decays of the resonances, $R \to \phi B$. These LECs appear in $\mL^{(1)}_{\phi R_{-}B}$, $ \mL^{(1)}_{\phi R_{+}B}$, displayed in Eqs.~(\ref{eq:L1phiRB}) and (\ref{eq:L1phiB*B}), respectively. 

In general, we follow the spirit of \cite{Borasoy:1999md} to determine these strong resonance LECs. 
However, for the $1/2^+$ excited octet, we update these values by performing a fit to the strong decays, given the significant improvement of data precision in recent years. We consider the six decays $N(1440)\to N \pi$, $\Lambda(1600)\to N K$, $\Lambda(1600)\to \Sigma \pi$, $\Sigma(1660)\to N K$, $\Sigma(1660)\to \Lambda \pi$ and $\Sigma(1660)\to \Sigma \pi$. 
The respective decay width is computed from the Lagrangian of Eq.~\eqref{eq:L1phiB*B} and fitted to data \cite{CBELSATAPS:2015kka,Sarantsev:2019xxm}, resulting in a reasonably good description. Details are provided in Appendix \ref{app:det-strong-Roper}. This constitutes an update of the previous $D^*$ and $F^*$ constraints from Ref.~\cite{Borasoy:1996bx}, as listed in Table~\ref{tab:new-fit-DF}.

\begin{table}[ht]
    \centering
    $\begin{array}{ccc} 
           {\rm \bf{LEC}} & D^* & F^* \\ \hline
            {\rm this \ work} & 0.713 \pm 0.107 & 0.342 \pm 0.051 \\
            \text{Ref.\cite{Borasoy:1996bx}} & 0.60 \pm 0.41 & 0.11\pm 0.41 \\
            \hline
        \end{array} $
    \caption{New value estimate for the strong resonance LECs $D^*, \ F^*$, with an assigned nominal error of 15\%.}
    \label{tab:new-fit-DF}
\end{table}

The situation for the states of the $1/2^-$ multiplet is more complicated, both from the experimental and from the theoretical point of view. The decay widths \cite{ParticleDataGroup:2024cfk} exhibit a significant flavor breaking, not accounted for when using (\ref{eq:L1phiRB}) at tree level. In fact, the nature of the lowest-lying $1/2^-$ states is extensively discussed in the literature.\footnote{This discussion is tied to our previous mentioning of the nature of the $\Lambda$ singlet state(s).} Certainly these states exhibit a strong admixture to meson-baryon configurations \cite{Kaiser:1995cy,Oller:2000ma,Garcia-Recio:2003ejq,Kolomeitsev:2003kt,Jido:2007sm,Bruns:2010sv,Doring:2025sgb}. In studies of the chiral structure of hadrons and of chiral restoration, the $N(1535)$ is discussed as a chiral partner of the nucleon \cite{Detar:1988kn,Jido:2001nt,Kummer:2025kch}. Of course, assumptions about the microscopic structure of a state crucially influence the coupling strengths to the various decay channels. In view of these complications we content ourselves with the employment of the state-of-the-art LEC values extracted in Ref.~\cite{Borasoy:1999md}. 
This determines $s_d=0.17$ and $s_f=-0.12$, to which we assigned an estimated error of 15\%.

\subsubsection{Contributions to s- and p-wave amplitudes}
\label{sec:diagrams}
\begin{figure}[ht!]
   \begin{subfigure}{0.13\textwidth}
       \begin{tikzpicture} 
    \begin{feynman}
        \vertex (pi) {}; 
        \vertex[right=1cm of pi, square dot] (v) {};
        \vertex[above=1.5cm of v] (v3);
        \vertex[above right=1.5cm of v] (pif) {};
        \vertex[right=1cm of v] (pf) {};
    \diagram* { 
    {[edge=thick](pi) -- (v) -- (pf), (v) -- [scalar] (pif)}, 
    };
    \end{feynman}
    \end{tikzpicture} 
       \caption{}
       \label{1a}
   \end{subfigure}
   \begin{subfigure}{0.19\textwidth} 
\includegraphics[width=\textwidth]{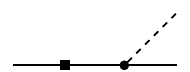}
       \caption{}
       \label{1b}
   \end{subfigure}
   \begin{subfigure}{0.19\textwidth} 
\includegraphics[width=\textwidth]{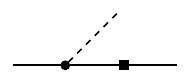}
       \caption{}
       \label{1c}
   \end{subfigure}
   \begin{subfigure}{0.19\textwidth} 
\includegraphics[width=\textwidth]{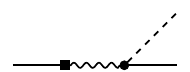}
       \caption{}
       \label{1d}
   \end{subfigure}
   \begin{subfigure}{0.19\textwidth} 
\includegraphics[width=\textwidth]{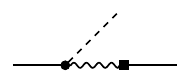}
       \caption{}
       \label{1e}
   \end{subfigure}
   \caption{Tree-level diagrams including resonances. The dashed, solid and wiggle lines correspond to meson, ground-state octet baryon ($B$) and $1/2^{\pm}$ baryon resonances ($R_{+}, \ R_{-} $), respectively. The box depicts the weak interaction mediated by the LECs $h_{D,F}$ and $w_{d,f}$ ($BR_{-}$), $d^*, f^*$ ($BR_{+}$) for the LO (\ref{eq:weaklag}) and resonance (\ref{eq:weak-s-res-L}, \ref{eq:weak-p-res-L}) Lagrangians, respectively. Diagrams (d) and (e) for $R_-$ and diagram (a) contribute to the {\it s}-wave amplitude. Diagrams (d) and (e) for $R_+$ together with diagrams (b) and (c) contribute to the {\it p}-wave.}
     \label{fig:tree-diags}
\end{figure}

\begin{figure}[ht!]
   \begin{subfigure}[b]{0.13\textwidth}
\includegraphics[width=\textwidth]{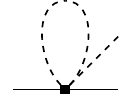}
       \caption{}
       \label{2a}
   \end{subfigure}
   \begin{subfigure}[b]{0.19\textwidth}
\includegraphics[width=\textwidth]{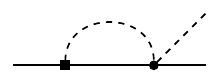}
       \caption{}
       \label{2b}
   \end{subfigure}
   \begin{subfigure}[b]{0.19\textwidth}
\includegraphics[width=\textwidth]{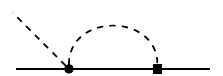}
       \caption{}
       \label{2c}
   \end{subfigure}
   \begin{subfigure}[t]{0.19\textwidth}
\includegraphics[width=\textwidth]{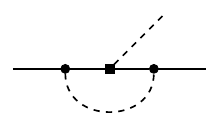}
       \caption{}
       \label{2d}
   \end{subfigure}
   \begin{subfigure}[t]{0.19\textwidth}
\includegraphics[width=\textwidth]{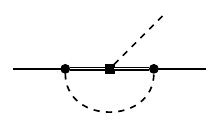}
       \caption{}
       \label{2e}
   \end{subfigure}
   \caption{Loop diagrams contributing to the {\it s}-wave amplitude. The types of lines and LECs are the same as in Fig.~\ref{fig:tree-diags}, except for the double line, representing states from the baryon decuplet ($T$), mediated by $h_C$ from (\ref{eq:weaklag}).}
     \label{fig:s-wave-diags}
\end{figure}

\begin{figure}[ht!]
   \begin{subfigure}{\myfigwidth}
\includegraphics[width=\textwidth]{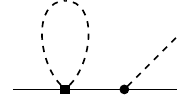}
       \caption{}
       \label{3a}
   \end{subfigure}
   \begin{subfigure}{\myfigwidth}
\includegraphics[width=\textwidth]{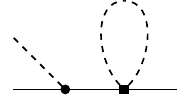}
       \caption{}
       \label{3b}
   \end{subfigure}
   \begin{subfigure}{\myfigwidth}
\includegraphics[width=\textwidth]{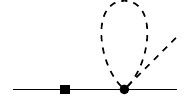}
       \caption{}
       \label{3c}
   \end{subfigure}
   \begin{subfigure}{\myfigwidth}
\includegraphics[width=\textwidth]{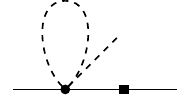}
       \caption{}
       \label{3d}
   \end{subfigure}
   
   \begin{subfigure}[b]{0.19\textwidth}
\includegraphics[width=\textwidth]{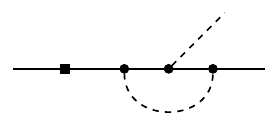}
       \caption{}
       \label{3e}
   \end{subfigure}
   \begin{subfigure}{0.19\textwidth}
\includegraphics[width=\textwidth]{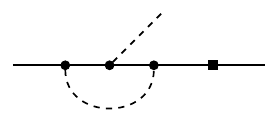}
       \caption{}
       \label{3f}
   \end{subfigure}
   \begin{subfigure}{0.19\textwidth}
\includegraphics[width=\textwidth]{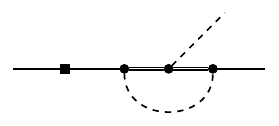}
       \caption{}
       \label{3g}
   \end{subfigure}
   \begin{subfigure}{0.19\textwidth}
\includegraphics[width=\textwidth]{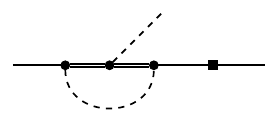}
       \caption{}
       \label{3h}
   \end{subfigure}
   \begin{subfigure}{0.19\textwidth}
\includegraphics[width=\textwidth]{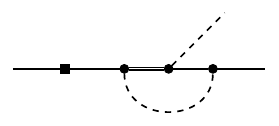}
       \caption{}
       \label{3i}
   \end{subfigure}
   \begin{subfigure}[t]{0.19\textwidth}
\includegraphics[width=\textwidth]{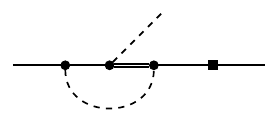}
       \caption{}
       \label{3j}
   \end{subfigure}
   \begin{subfigure}[t]{0.19\textwidth}
\includegraphics[width=\textwidth]{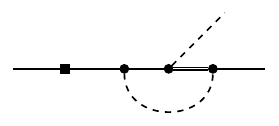}
       \caption{}
       \label{3k}
   \end{subfigure}
   \begin{subfigure}[t]{0.19\textwidth}
    \includegraphics[width=\textwidth]{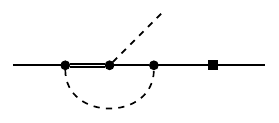}
       \caption{}
       \label{3l}
   \end{subfigure}
   \begin{subfigure}{0.19\textwidth}
\includegraphics[width=\textwidth]{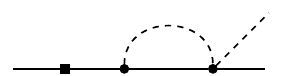}
       \caption{}
       \label{3m}
   \end{subfigure}
   \begin{subfigure}{0.19\textwidth}
       \includegraphics[width=\textwidth]{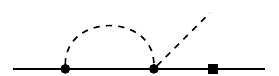}
       \caption{}
       \label{3n}
   \end{subfigure}
   \begin{subfigure}{0.19\linewidth} 
   \resizebox{\textwidth}{!}{%
    \begin{tikzpicture}
        \begin{feynman}
            \vertex (pi) {}; 
            \vertex[right=1cm of pi, square dot] (v) {};
            \vertex[right=1cm of v, dot] (v1) {};
            \vertex[above left=1.5cm of v1] (pif) {};
            \vertex[right=1.5cm of v1, dot] (v2) {};
            \vertex[right=1cm of v2] (pf) {};
        \diagram* { 
        {[edge=thick](pi) -- (v) -- (v1) -- (v2) -- (pf), (v1) -- [scalar] (pif), (v1) -- [scalar,half left] (v2)}, 
        };
        \end{feynman}
    \end{tikzpicture} }
       \caption{}
       \label{3o}
   \end{subfigure}
   \begin{subfigure}{0.19\textwidth}
   \resizebox{\textwidth}{!}{%
       \begin{tikzpicture}
            \begin{feynman}
                \vertex (pi) {}; 
                \vertex[right=1cm of pi, dot] (v1) {};
                \vertex[above left=1.5cm of v1] (pif) {};
                \vertex[right=1.5cm of v1, dot] (v2) {};
                \vertex[right=1cm of v2, square dot] (v) {};
                \vertex[right=1cm of v] (pf) {};
            \diagram* { 
            {[edge=thick](pi) -- (v1) -- (v2) -- (v) -- (pf), (v1) -- [scalar] (pif), (v1) -- [scalar,half left] (v2)}, 
            };
            \end{feynman}
        \end{tikzpicture} } 
       \caption{}
       \label{3p}
   \end{subfigure}
   \begin{subfigure}{0.2\linewidth} 
   \resizebox{\textwidth}{!}{%
    \begin{tikzpicture}
        \begin{feynman}
            \vertex (pi) {}; 
            \vertex[right=1cm of pi, dot] (v1) {};
            \vertex[right=0.75cm of v1, square dot] (v) {};
            \vertex[right=0.75cm of v, dot] (v2) {};
            \vertex[right=1.cm of v2, dot] (v3) {};
            \vertex[above right=1.5cm of v3] (pif) {};
            \vertex[right=1cm of v3] (pf) {};
        \diagram* { 
        {[edge=thick](pi) -- (v1) -- (v) -- (v2) --(v3) -- (pf), (v3) -- [scalar] (pif), (v1) -- [scalar,half left] (v2)}, 
        };
        \end{feynman}
    \end{tikzpicture} }
       \caption{}
       \label{3q}
   \end{subfigure}
   \begin{subfigure}{0.2\linewidth}
   \resizebox{\textwidth}{!}{%
    \begin{tikzpicture}
        \begin{feynman}
            \vertex (pi) {}; 
            \vertex[right=1cm of pi, dot] (v1) {};
            \vertex[above left=1.5cm of v1] (pif) {};
            \vertex[right=1.cm of v1, dot] (v2) {};
            \vertex[right=0.75cm of v2, square dot] (v) {};
            \vertex[right=0.75cm of v, dot] (v3) {};
            \vertex[right=1cm of v3] (pf) {};
        \diagram* { 
        {[edge=thick](pi) -- (v1) -- (v2) -- (v) --(v3) -- (pf), (v1) -- [scalar] (pif), (v2) -- [scalar,half left] (v3)}, 
        };
        \end{feynman}
    \end{tikzpicture} }
       \caption{}
       \label{3r}
   \end{subfigure}%
   \begin{subfigure}{0.2\linewidth} 
   \resizebox{\textwidth}{!}{%
    \begin{tikzpicture}
        \begin{feynman}
            \vertex (pi) {}; 
            \vertex[right=1cm of pi, dot] (v1) {};
            \vertex[right=0.75cm of v1, square dot] (v) {};
            \vertex[right=0.75cm of v, dot] (v2) {};
            \vertex[right=1.cm of v2, dot] (v3) {};
            \vertex[above right=1.5cm of v3] (pif) {};
            \vertex[right=1cm of v3] (pf) {};
        \diagram* { 
        {[edge=thick](pi) -- (v1) -- [double] (v) -- [double] (v2) --(v3) -- (pf), (v3) -- [scalar] (pif), (v1) -- [scalar,half left] (v2)}, 
        };
        \end{feynman}
    \end{tikzpicture} }
       \caption{}
       \label{3s}
   \end{subfigure}
   \begin{subfigure}{0.2\linewidth} 
   \resizebox{\textwidth}{!}{%
    \begin{tikzpicture}
        \begin{feynman}
            \vertex (pi) {}; 
            \vertex[right=1cm of pi, dot] (v1) {};
            \vertex[above left=1.5cm of v1] (pif) {};
            \vertex[right=1.cm of v1, dot] (v2) {};
            \vertex[right=0.75cm of v2, square dot] (v) {};
            \vertex[right=0.75cm of v, dot] (v3) {};
            \vertex[right=1cm of v3] (pf) {};
        \diagram* { 
        {[edge=thick](pi) -- (v1) -- (v2) -- [double] (v) -- [double] (v3) -- (pf), (v1) -- [scalar] (pif), (v2) -- [scalar,half left] (v3)}, 
        };
        \end{feynman}
    \end{tikzpicture} }
       \caption{}
       \label{3t}
   \end{subfigure}
   \begin{subfigure}{0.16\textwidth}
\includegraphics[width=\textwidth]{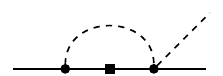}
       \caption{}
       \label{3u}
   \end{subfigure}
   \begin{subfigure}{0.16\textwidth}
\includegraphics[width=\textwidth]{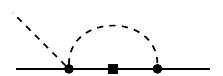}
       \caption{}
       \label{3v}
   \end{subfigure}
   \begin{subfigure}{0.16\textwidth}
\includegraphics[width=\textwidth]{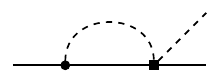}
       \caption{}
       \label{3w}
   \end{subfigure}
   \begin{subfigure}{0.16\textwidth}
\includegraphics[width=\textwidth]{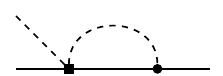}
       \caption{}
       \label{3x}
   \end{subfigure}
   \caption{Loop diagrams that contribute to the {\it p}-wave amplitude. The types of lines are the same as in Figs.~\ref{fig:tree-diags}-\ref{fig:s-wave-diags}.}
     \label{fig:p-wave-diags}
\end{figure}

The tree-level diagrams (including resonances) are depicted in Fig.~\ref{fig:tree-diags}. The loop diagrams contributing to the {\it s-} and {\it p-}waves are depicted in Fig.~\ref{fig:s-wave-diags} and~\ref{fig:p-wave-diags}, respectively. Notice that all {\it p}-wave diagrams in Fig.~\ref{fig:p-wave-diags}, except \ref{3u}-\ref{3x}, contain the direct transition, i.e.\ the mixing of two baryon states. Some of the topologies have not been included in previous analyses \cite{Jenkins:1991bt, Borasoy:1998ku, AbdEl-Hady:1999llb}. In particular, there are diagrams (\ref{3m}-\ref{3p}, \ref{3u}-\ref{3x}) that vanish in the heavy-baryon approximation where the baryons do not propagate spin information. This is different in the relativistic calculation. Contributions at order $\mO(M_K^2)$ from such diagrams are not excluded in the EOMS scheme. We recall that the wave function renormalization is not represented diagrammatically, but is included in our calculations. Before proceeding to compare with data, the theoretical uncertainty is assessed in the next section.

\subsubsection{Estimation of the theoretical uncertainty}
\label{sect:uncertainties}
As in any effective field theory computation, the effect of the systematic uncertainty has to be assessed. We quantify a theoretical error and incorporate it in the denominator of the $\chi^2$ in order to make a meaningful comparison between data and theory. On the one hand, the truncation of the chiral series at NLO introduces an uncertainty, which is estimated to be of the size of the first missing order $\mO(M_K^3/\Lambda_{\chi}^3)$ (see for instance the detailed discussion in Ref.~\cite{Epelbaum:2014efa}). Here $\Lambda_{\chi}\simeq 1$ GeV is the scale of spontaneous symmetry breaking of QCD. For a given weak decay amplitude $\mM$, the error associated to the truncation of the chiral series is approximated by $\sigma(\mM_{\rm trunc})=(M_K^3/\Lambda_{\chi}^3)\ \mM\simeq 0.15\ \mM_{\rm exp}$, where $\mM_{\rm exp}$ is the experimental value of the amplitude. On the other hand, the LECs related to the strong decays have an estimated error of 15\%. Therefore, we propagate this uncertainty to the amplitude, denoting it $\sigma(\mM_{\rm res})$. 
As a result, the theoretical error is given by
\begin{equation}\label{eq:theo_err}
    \sigma(\mM_{\rm theory})^2=\sigma(\mM_{\rm trunc})^2+\sigma(\mM_{\rm res})^2 \ .
\end{equation}
This theoretical uncertainty dominates over the experimental one and the two of them are summed in quadrature in the $\chi^2$ when performing the fit\footnote{In order to evaluate $\sigma(\mM_{\rm res})$, the values of $\omega_d$, $\omega_f$, $d^*$ and $f^*$ are first taken from a fit to the weak hyperon decay amplitudes in which $\sigma(\mM_{\rm res})$ is set to zero. The fit is then repeated updating $\sigma(\mM_{\rm theory})$, hence the $\chi^2$, with the newly extracted $\sigma(\mM_{\rm res})$, until convergence is reached.}. This allows to determine amplitudes and LECs and at the same time reflecting the degree of accuracy of the theory. Moreover, if the theoretical uncertainty is neglected, the highly-precise experimental data points have an unrealistically large impact on the obtained LECs. Therefore, both experimental and theoretical errors are considered in the $\chi^2$ that is then minimized and reported in the tables. The errors of the fitted LECs that we quote correspond to the $1\sigma$ deviation from the $\chi^2$ minimum. The reported results of the theoretical determination of the decay amplitudes have error $\sigma(\mM_{\rm theory})$ given in Eq.~\eqref{eq:theo_err}.
\section{Fit results}
\label{sec:fit}

\subsection{ChPT without octet resonances}
\label{sec:ChPT-no-res}

In line with Ref.\ \cite{Borasoy:1999md} we regard the resonances as important players. But we should check how far we can come without them. After all, an important new aspect of the present work is the fact that our ChPT amplitudes are determined in the relativistic EOMS framework and not in the traditional heavy-baryon approach. For the calculation of electromagnetic form factors it turned out that this difference matters a lot \cite{Geng:2008mf}.

Therefore, a combined fit of the relativistic pure ChPT amplitudes of {\it s-} and {\it p-}waves is performed. In this first approach, the theory contains the decuplet fields explicitly but no NLO counterterms. The input values for the strong ChPT LECs are given in Table~\ref{tab:strong_LECs}. We stress that those LECs are not fitted to the weak decays. Only the weak LECs from (\ref{eq:weaklag}) are regarded as free parameters in our fits.
\begin{table}[ht]
    \centering
    \begin{tabular}{cc}
        \textbf{LEC} & \textbf{Ref.} \\ \toprule
        $\mathcal{C}=1.6$ & \cite{Granados:2017cib} \\
        $\mathcal{H}=-1.9\pm0.7$ & \cite{Jenkins:1991bt} \\
        $D=0.8$ & \cite{Granados:2017cib} \\
        $F=0.46$ & \cite{Granados:2017cib} \\
    \end{tabular}
    \caption{Values of ChPT LECs appearing in $ \mL^{(1)}_{\phi BT}$, Eq.~(\ref{eq:stronglag}). Note that the value of the decuplet-decuplet-meson coupling constant $\mathcal{H}$ is not very well known \cite{Bertilsson:2023htb}.}
    \label{tab:strong_LECs}
\end{table}

Such a parametrization is unable to describe simultaneously the {\it s-} and {\it p}-wave data, as apparent from the results in Tab.~\ref{tab:iso-fit-comb-nores}\footnote{In addition, we also conducted a fit excluding the decuplet fields. It is not displayed here, because it does not improve the reduced $\chi^2$.} (see Sect.~\ref{sect:uncertainties} for details on the uncertainties). In the table, the values of the different orders show that the convergence of the chiral series is in general slow. 
\begin{table}[ht]
        \centering 
        $\renewcommand{\arraystretch}{1.4}
        \begin{array}{ccccccccc} 
        \hline 
           {\rm \bf{LEC}} \ [G_F m_{\pi^+}^2\sqrt{2}F_\pi] & h_D & h_F & h_C \\ 
            {\rm \bf{Value}} & -0.359 \pm 0.056 & 0.205 \pm 0.028 & -0.448 \pm 0.087 \\
            \hline
        \end{array} $
\\
        \begin{tabular}{c}
        $\tilde{\chi}^2\equiv\chi^2/(n_{\rm dat}-n_{\rm pars})=240/(8-3)=48$ \\
        \end{tabular}
        $\begin{array}{rrrrrrrrrrr}
        \hline
            &   s_{\rm expt} &   s_{\rm theory} &   s_{\rm tree} &    \Delta s_{\rm 8} &   \Delta s_{\rm 10} &   p_{\rm expt} & p_{\rm theory} & p_{\rm tree} & \Delta p_{\rm 8} & \Delta p_{\rm 10} \\
        \toprule
          L_{11}^\Sigma  &   -0.936 &     -0.172 &   -0.278 &  0.100 &  0.006 &    2.714 &     -0.034 &   -0.595 & -0.091  &  0.652 \\ 
          L_{13}^\Sigma   &    1.948 &      0.079 &    0.554 & -0.473 & -0.002 &    0.015 &      0.015 &    0.302 & -0.723 &  0.436 \\
          L_{11}^\Lambda  &    1.412 &     -0.088 &    0.104 & -0.072 & -0.119 &    0.625  &      0.408 &    0.078 & -0.072 &  0.402 \\
          L_{12}^\Xi     &   -2.046 &      0.534  &   -0.398 &  0.491 &  0.441 &    0.397 &      0.409  &   -0.217 &  0.596 &  0.031 \\
        \hline
        \end{array}$   
        $\begin{array}{lrrrrrrrrrr}
        \hline
            &   s_{\rm expt} &   s_{\rm theory} &   s_{\rm tree} &     \Delta s_{8} &   \Delta s_{10} &   p_{\rm expt} & p_{\rm theory} & p_{\rm tree} & \Delta p_{8} & \Delta p_{10} \\
         \toprule
         \Sigma^+\to n\pi^+     &    0.023 &     -0.094 &    0     & -0.094 &  0     &    1.815 &     -0.003 &   -0.295 & -0.277 &  0.569 \\
         \Sigma^+\to p\pi^0     &   -1.337 &     -0.114 &   -0.399 &  0.276 &  0.008 &    1.272 &     -0.044 &   -0.425 &  0.264 &  0.117 \\
         \Sigma^-\to n\pi^-     &    1.914 &      0.081  &   0.564 & -0.482 & -0.002 &    0.015 &      0.015 &    0.303 & -0.725  &  0.437 \\
         \Lambda\to p\pi^-      &    1.403 &     -0.088 &    0.104 & -0.073 & -0.120 &    0.625  &      0.408 &    0.078 & -0.072 &  0.402 \\
         \Lambda\to n\pi^0      &   -0.992 &      0.060 &   -0.074 &  0.051  &  0.083 &   -0.442 &     -0.303   &   -0.057 &  0.049 & -0.295 \\
         \Xi^-\to \Lambda\pi^-  &   -2.046 &      0.534  &   -0.398 &  0.491 &  0.441 &    0.397 &      0.409  &   -0.217 &  0.596 &  0.031 \\
         \Xi^0\to \Lambda\pi^0  &    1.447 &     -0.380 &    0.281  & -0.349 & -0.313  &   -0.281 &     -0.293 &    0.155 & -0.421 & -0.028 \\
        \hline
    \end{array}$ 
    \caption{Results for the physical and isospin amplitudes from the combined fit to the {\it s}- and {\it p}-waves in relativistic ChPT without the contribution of the resonant octets. Loop contributions from octet (index 8) and decuplet (index 10) are displayed separately. The LECs are given in units of $G_F m_{\pi^+}^2$, as well as of the physical pion decay constant $F_\pi = 92.4\times10^{-3}\,$GeV.} 
    \label{tab:iso-fit-comb-nores}
    \end{table}

 We conclude that a satisfactory combined description of both partial waves is not achieved by relativistic ChPT, if one drops all counterterms. This fact motivates our final step where we include the resonances as a way to estimate the neglected NLO counterterms.
    
\subsection{ChPT including octet resonances}
\label{sec:ChPT-res}

In total, we have explored two different fit strategies. First, a separate fit to {\it s}- and {\it p}-waves; second, a combined fit. In the following, we present details on the second fit, but comment only briefly on the first option. 

If one fits to the two partial waves separately, one can obtain a good agreement with the experimental data. But it leads to the problem that the fit parameters are not sufficiently constrained. Depending on the details one can produce incompatible values for the common LECs $h_D$, $h_F$, $h_C$. Otherwise the combined fit to both partial waves shows the same qualitative features as the separate fit. Therefore we focus on the combined fit being more constraining. 
    
\begin{figure}[ht]
        \centering
        \includegraphics[width=\linewidth]{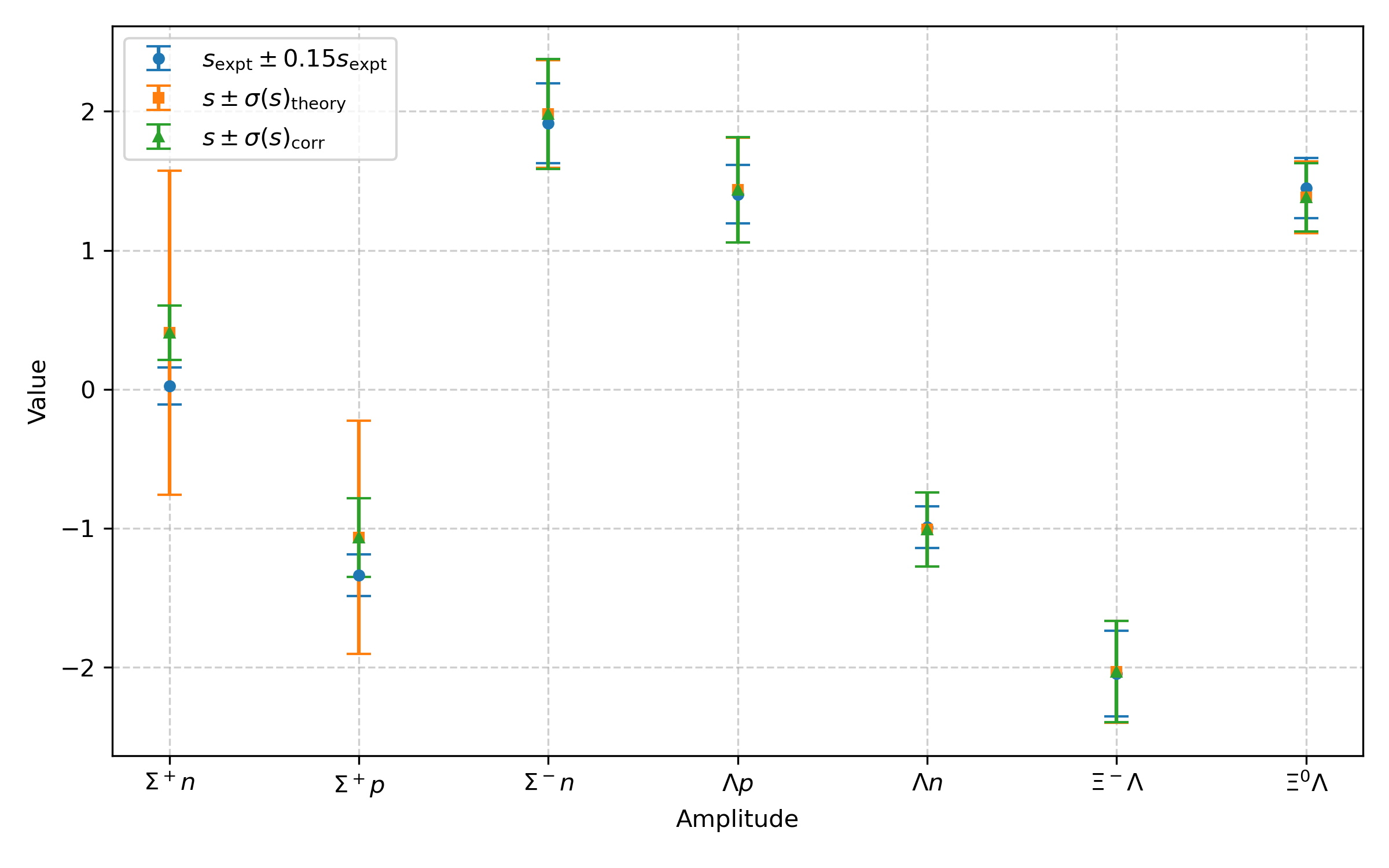}
        \includegraphics[width=\linewidth]{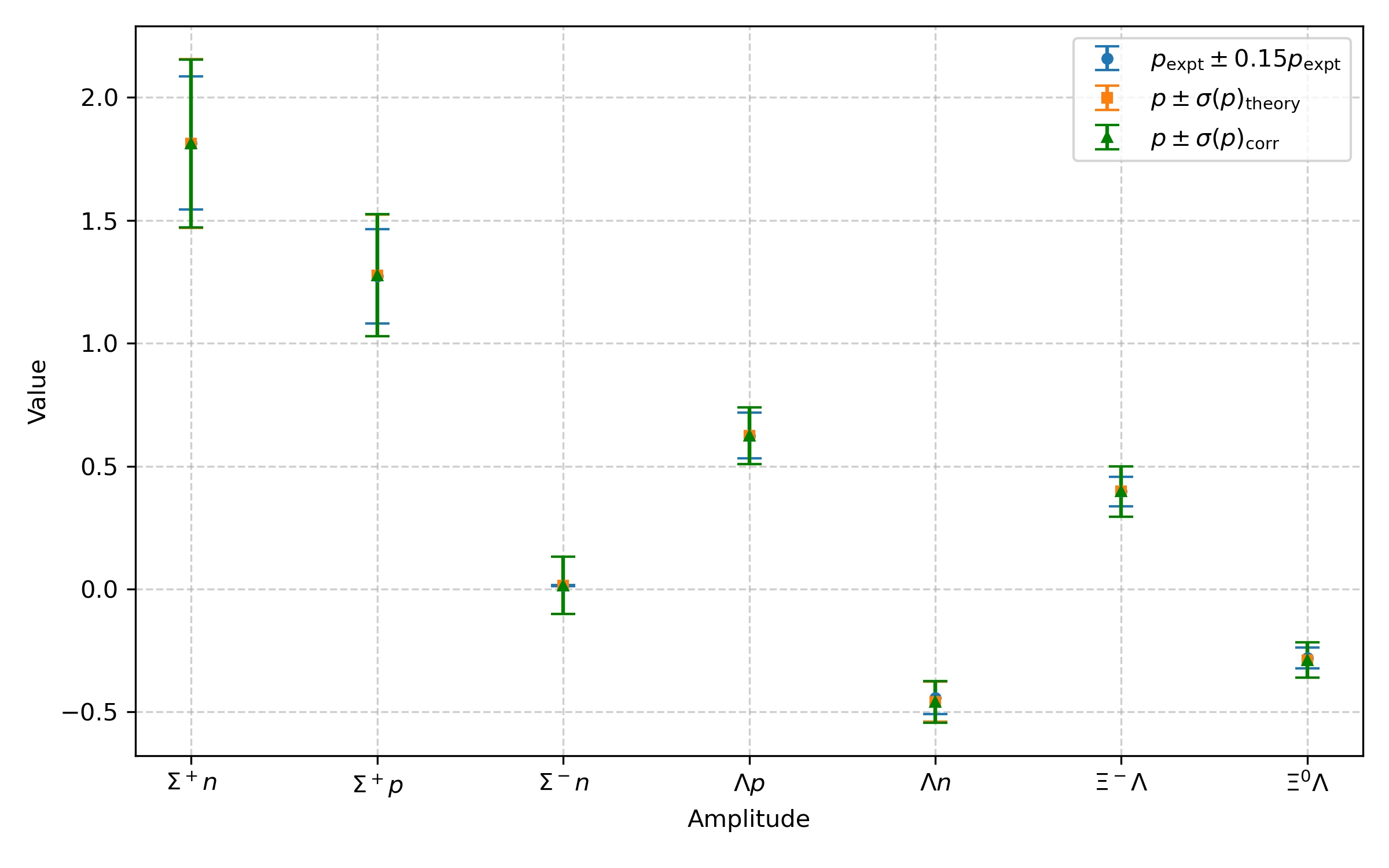}
        \caption{Results of the combined fit to the {\it s}- and {\it p}-wave amplitudes via the isospin amplitudes $L_{2\Delta I, 2I}$. The theoretical error, $\sigma(\mM)_{\rm theory}$, is defined in Eq.~\eqref{eq:theo_err}. The error $\sigma(\mM)_{\rm corr}$ corresponds to the 1$\sigma$ error propagated from the fitted LECs including the correlation coefficients and the uncertainty in the LEC $\mathcal{H}$ given in Table \ref{tab:strong_LECs}. 
        }
        \label{fig:iso-fit-comb-res}
    \end{figure}

\begin{figure}[ht]
        \centering
        \includegraphics[width=\linewidth]{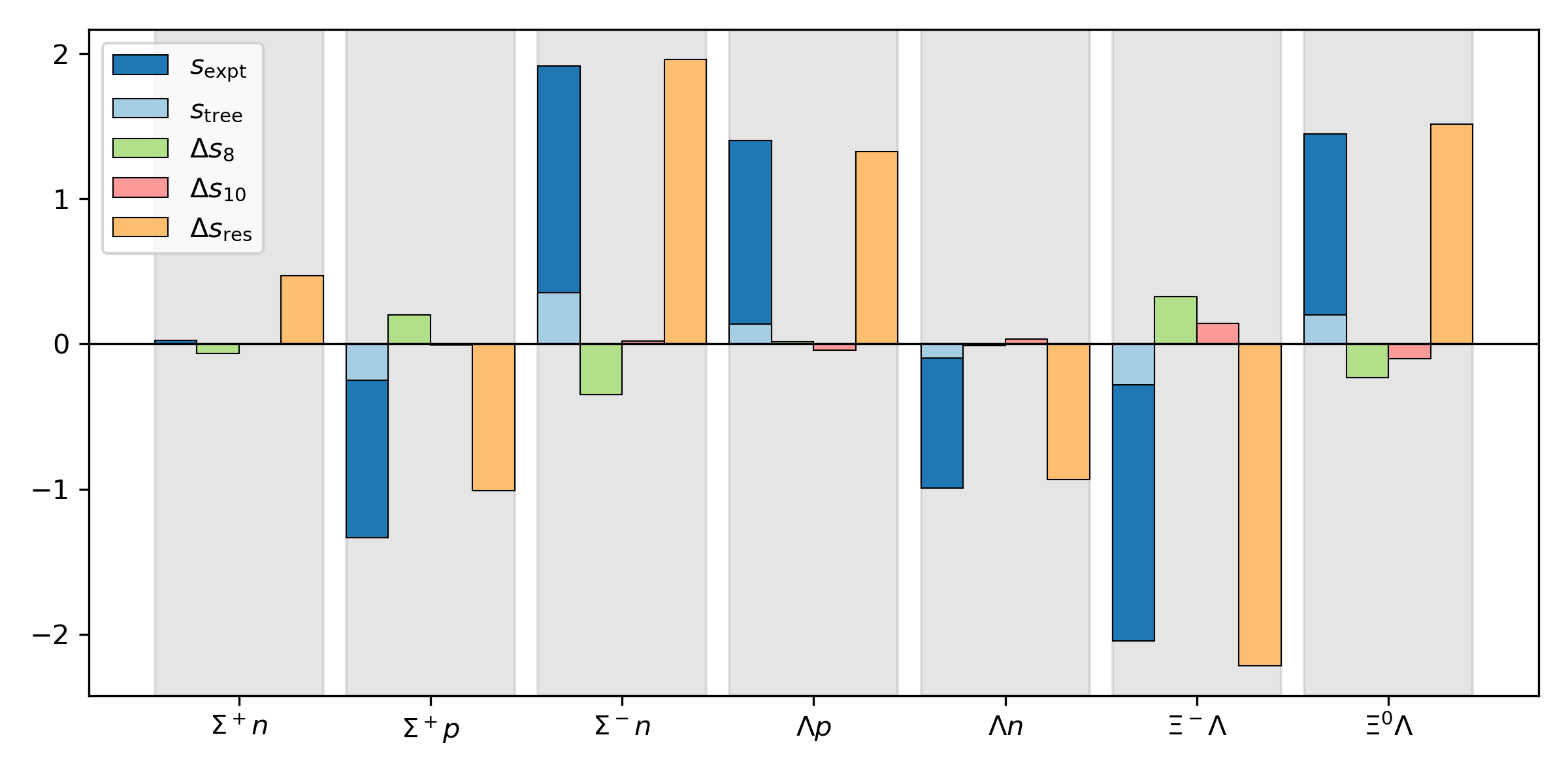}
        \includegraphics[width=\linewidth]{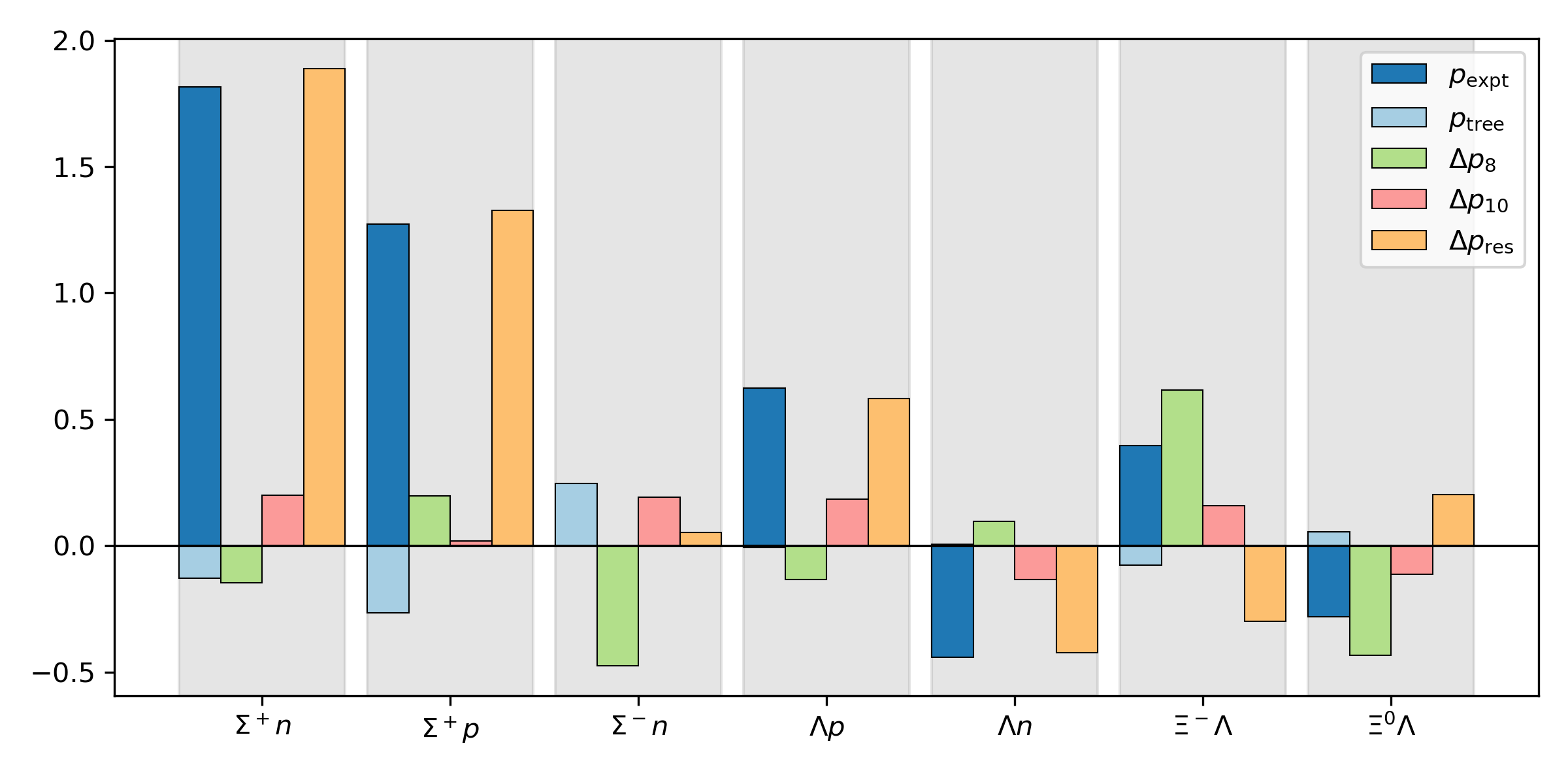}
        \caption{Size comparison of $l_{\rm expt}$ to the various LO and NLO contributions to the physical amplitudes, corresponding to the values in the bottom half of Table \ref{tab:iso-fit-comb-res}.}
        \label{fig:iso-fit-res-components}
    \end{figure}

\begin{figure}
        \centering
        \includegraphics[width=\linewidth]{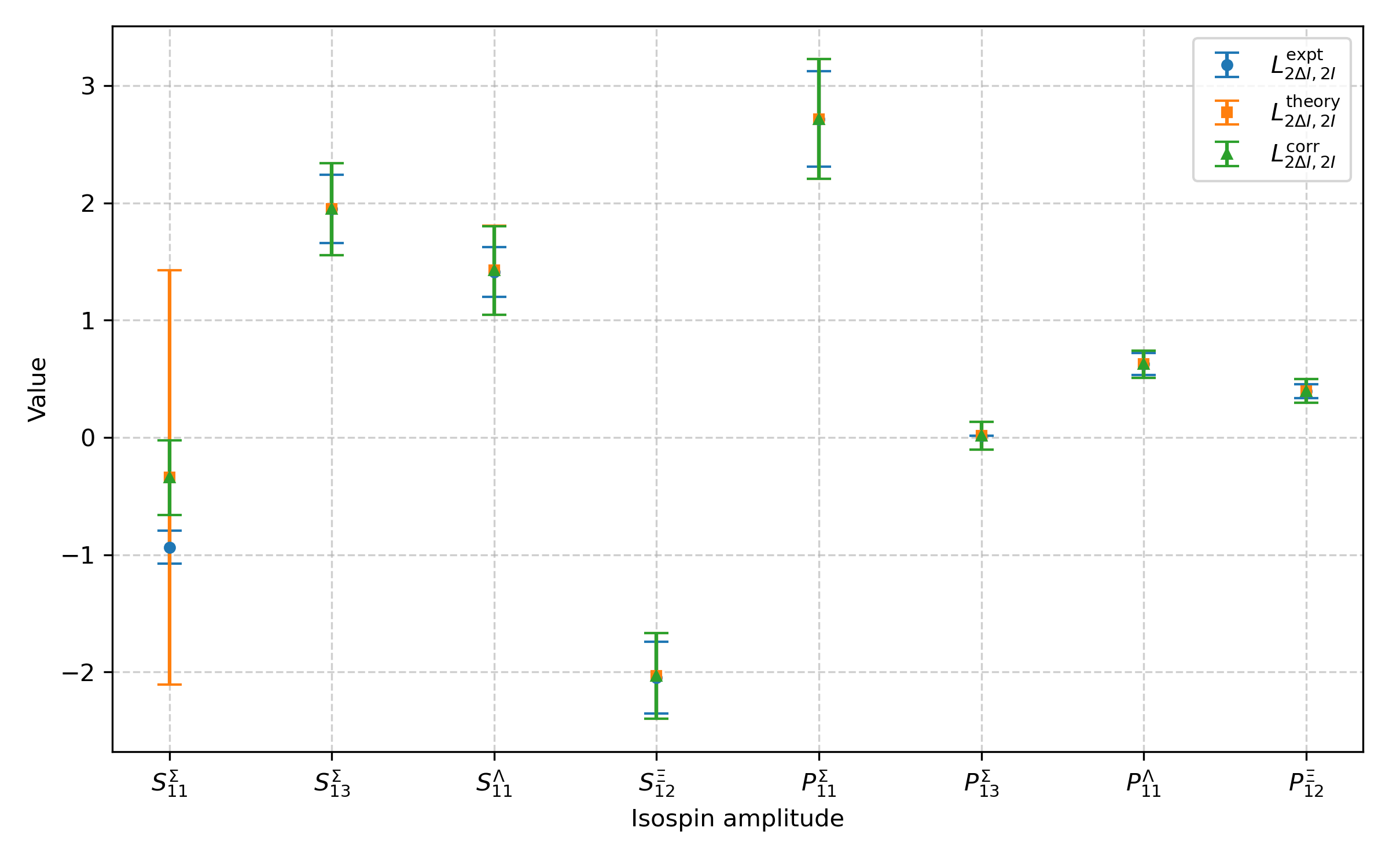}
        \caption{Results of the combined fit to the {\it s}- and {\it p}-waves for the isospin amplitudes $L_{2\Delta I, 2I}$. See also Fig.\ \ref{fig:iso-fit-comb-res} and Sec.\ \ref{sect:uncertainties} for details about the uncertainties.}
        \label{tab:SPiso-fit-comb-res} 
        \end{figure}

\begin{figure}[ht]
        \centering
        \includegraphics[width=\linewidth]{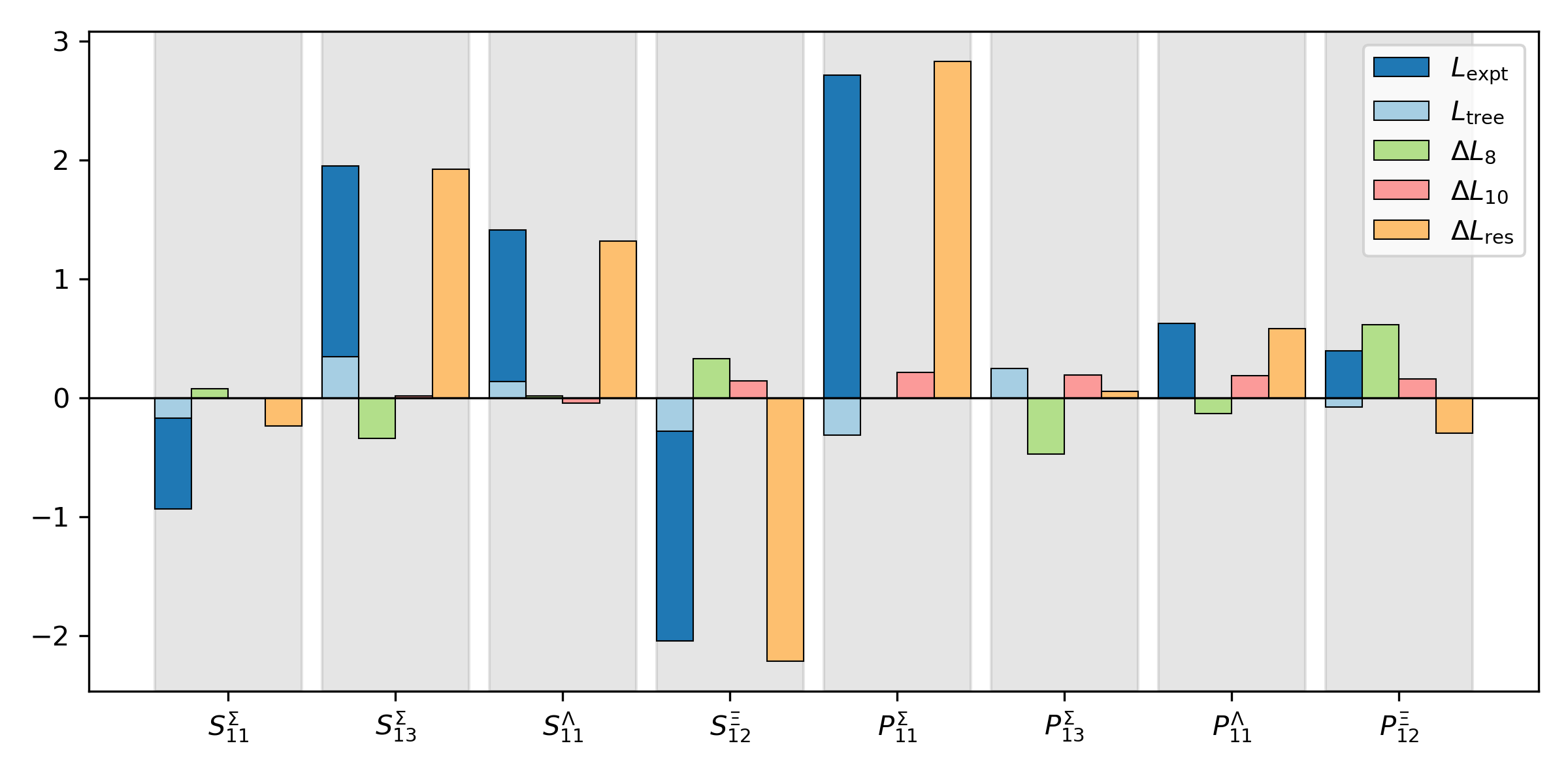}
        \caption{Size comparison of $L_{\rm expt}$ to the various LO and NLO contributions to the isospin amplitudes, corresponding to the values in the top half of Table \ref{tab:iso-fit-comb-res}.}
        \label{fig:SPiso-fit-res-components}
    \end{figure}

The main result of this work consists in the combined fit of {\it s-} and {\it p-}waves, including the $\Delta I=1/2$ information from all seven decays. The results are displayed in Table~\ref{tab:iso-fit-comb-res} and Figs.\ \ref{fig:iso-fit-comb-res}-\ref{fig:SPiso-fit-res-components} (see Sect.~\ref{sect:uncertainties} for details on the uncertainties). 
As one can see in Fig.~\ref{fig:iso-fit-comb-res}, the theory is able to describe the data of both partial waves simultaneously.      

We would like to point out some details about the fit results. The {\it s}-wave decays for the $\Sigma^{+}$ channels, $\Sigma^{+}\to n \pi^{+}$ and $\Sigma^{+}\to p \pi^{0}$, have a relatively large uncertainty due to a cancellation between the contributions of two LECs associated to the strong decays of $1/2^{-}$ resonances, namely $s_d$ and $s_f$. The low value of the reduced $\chi^2$ could be due to a too conservative estimation of the theoretical uncertainty. Moreover, the LECs $h_{D}$, $h_C$, and $d^*$ are not well constrained by our fit. We stress that the goal of this project is not a high-precision estimate of the weak LECs. Instead it constitutes a systematic investigation of whether it is possible to reproduce the experimental data, if one combines a relativistic EOMS calculation in ChPT with a resonance-saturation estimate of the LECs at NLO. 
 
Irrespective of choosing separate or a common fit for the two partial waves, it turns out that the contributions of the resonances dominate the amplitudes (Fig.~\ref{fig:iso-fit-res-components}). This result indicates that the resonances play an important role in non-leptonic hyperon decays, suggesting that such resonances should be accounted for in any computation of these decays. From a pure ChPT point of view, this implies that an NLO effect (the resonances are a proxy for the LECs at NLO) is more important than the LO contributions. This observed slow convergence of the ChPT expansion, also reflected by the sizeable theoretical uncertainty, suggests that a higher-order calculation would be of interest. In this direction, the aforementioned resonances would still be important. It might even be necessary to extend the ChPT framework and include the resonances as dynamical degrees of freedom. We will come back to this point in the concluding section. 

Finally, since the fit is performed using the eight independent amplitudes $L_{1,2I}$ extracted from data, we display their estimate from the fit in Fig.~\ref{tab:SPiso-fit-comb-res}. We also show the sizes of the different contributions to a given process in Fig.~\ref{fig:SPiso-fit-res-components}. The interplay between the slow convergence of ChPT and the resonance contributions is again evident.
    \begin{table}[ht]
        \centering 
        \resizebox{\textwidth}{!}{$
        \renewcommand{\arraystretch}{1.4}
        \begin{array}{ccccccccc} 
        \hline 
           {\rm \bf{LEC}} \ [G_F m_{\pi^+}^2\sqrt{2}F_\pi] & h_D & h_F & h_C & w_d & w_f & d^* & f^* \\ 
            {\rm \bf{Value}} & -0.181 \pm 0.252 & 0.171 \pm 0.041 & -0.090 \pm 0.472 &  -7.99 \pm 3.82 & 7.91 \pm 1.32 & 1.10 \pm 1.55 & -2.33 \pm 1.50  \\ 
            \hline
        \end{array} $
        }
        \\
        \centering
        $\begin{array}{c}
        \tilde{\chi}^2\equiv\chi^2/(n_{\rm dat}-n_{\rm pars})=0.11/(8-7)=0.11 
        \end{array} $
         \resizebox{\textwidth}{!}{$
         \begin{array}{lrrrrrrrrrrrr} \hline
             &  s_{\rm expt} & s_{\rm theory} &  s_{\rm tree} & \Delta s_{\rm 8} & \Delta s_{\rm 10} & \Delta s_{\rm res} & p_{\rm expt} & p_{\rm theory} &  p_{\rm tree} & \Delta p_{\rm 8} & \Delta p_{\rm 10} & \Delta p_{\rm res} \\ \toprule
         L_{11}^\Sigma    &  -0.936 &     -0.342 &   -0.173 &  0.076 & -0.006 &  -0.238 &    2.714 &      2.715 &   -0.317 & -0.008 &  0.213 &   2.828 \\
         L_{13}^\Sigma    &   1.948 &      1.947 &    0.346 & -0.341 &  0.018 &   1.924 &    0.015 &      0.015 &    0.246 & -0.475 &  0.192 &   0.053 \\
         L_{11}^\Lambda   &    1.412 &      1.424 &    0.135  &  0.015 & -0.044 &   1.319 &    0.625  &  0.624  &   -0.006 & -0.134 &  0.183 &   0.581 \\
         L_{12}^\Xi       &   -2.046 &     -2.032 &   -0.284 &  0.327 &  0.140  &  -2.216 &    0.397 &   0.397 &   -0.078 &  0.616 &  0.157 &  -0.298 \\
        \end{array} $
        }
        \resizebox{\textwidth}{!}{$
         \begin{array}{lrrrrrrrrrrrr}
         \hline
             &  s_{\rm expt} & s_{\rm theory} &  s_{\rm tree} & \Delta s_{8} & \Delta s_{10} & \Delta s_{\rm res} & p_{\rm expt} & p_{\rm theory} &  p_{\rm tree} & \Delta p_{8} & \Delta p_{10} & \Delta p_{\rm res} \\ \toprule
           \Sigma^+\to n\pi^+     &    0.023 &  0.407 &    0     &     -0.065 &    0     &    0.472 &   1.815 &    1.813 &     -0.129 &   -0.148 &    0.201 &   1.889 \\
           \Sigma^+\to p\pi^0     &   -1.337 & -1.066 &   -0.249  &      0.201 &   -0.008 &   -1.010 &   1.272 &    1.277 &     -0.265 &    0.197  &    0.017 &   1.328 \\
           \Sigma^-\to n\pi^-     &    1.914 &  1.981 &    0.352  &     -0.348 &    0.018 &    1.958 &   0.015 &    0.015 &      0.246 &   -0.476 &    0.192 &   0.053 \\
           \Lambda\to p\pi^-      &    1.403 &  1.434 &    0.136 &      0.015 &   -0.045 &    1.327 &   0.625  &    0.624  &     -0.006 &   -0.134 &    0.183 &   0.581 \\
           \Lambda\to n\pi^0      &   -0.992 & -1.009 &   -0.096 &     -0.011 &    0.031 &   -0.933 &  -0.442 &   -0.459 &      0.005 &    0.096 &   -0.135 &  -0.425 \\
           \Xi^-\to \Lambda\pi^-  &   -2.046 & -2.032 &   -0.284 &      0.327 &    0.140  &   -2.216 &   0.397 &    0.397 &     -0.078 &    0.616 &    0.157 &  -0.298 \\
           \Xi^0\to \Lambda\pi^0  &    1.447 &  1.382 &    0.200 &     -0.232 &   -0.101 &    1.515 &  -0.281 &   -0.289  &      0.056 &   -0.434 &   -0.113  &   0.202 \\
           \hline
        \end{array} $
        }      
        \caption{Results of the combined fit to {\it s}- and {\it p}-waves including tree-level diagrams with resonances. The total theory result (label ``theory'') is split up into the contributions from LO chPT (tree), the loop contributions with only octet baryons in the loop (8), the loop contributions involving decuplet states (10), and the contributions from resonances as a proxy for the NLO counterterms (res). The uncertainties of the theory results are shown in Figs.\ \ref{fig:iso-fit-comb-res} and \ref{tab:SPiso-fit-comb-res}.
        }
        \label{tab:iso-fit-comb-res} 
    \end{table}

\clearpage
\section{Conclusion and outlook} 
\label{sec:Conclusion-and-outlook}
In this work, the longstanding problem of hyperon non-leptonic decays is assessed in NLO relativistic Chiral Perturbation Theory with explicit decuplet fields. The NLO counterterms are approximated by the contributions of the integrated-out spin-$1/2^-$ and  spin-$1/2^+$ resonance octets. The first multiplet contributes to the $s$-wave and the latter to the $p$-wave. Such procedure introduces four additional unknown weak LECs on top of the three unknown weak ChPT constants. Moreover, a necessary ingredient are the LECs governing the strong decays of the resonances. In this regard we update the values of the Roper octet LECs by fitting to the results of Ref.~\cite{CBELSATAPS:2015kka,Sarantsev:2019xxm}, while the values in Ref.~\cite{Borasoy:1999md} are employed for the negative-parity multiplet.

 With respect to the data to which the ChPT amplitudes are compared, we emphasise the inclusion of recent experimental results from BESIII~\cite{BESIII:2018cnd,BESIII:2021ypr} that yield a significant change in the polarization asymmetry with regard to previous results. Moreover, the empirical data that constitute the basis of our analysis are extracted with minimum assumptions from an isospin decomposition and taking into account experimental final-state phase-shifts. 

First, we have shown that relativistic ChPT without NLO counterterms does not describe the $s$- and $p$-wave amplitudes together and the convergence of the chiral series is slow. Afterwards we have incorporated the contribution from the two integrated-out resonant octets as an approximation of such counterterms in the theory.  
With all parameters related to strong processes fixed from outside, we have in total seven weak LECs in our extended Lagrangian that were fitted to the non-leptonic hyperon decays.

Remarkably, both {\it s}- and {\it p}-waves can be described simultaneously by the theory. In general the convergence of the chiral series is still slow, which is reflected in our conservative uncertainty estimation and results in a not very tightly constrained fit. Expanding on the conclusions of Ref.\ \cite{Jenkins:1991bt}, the contribution of the spin-3/2 fields to the relativistic amplitudes is eclipsed by the relative weight of the resonance-exchange diagrams -- as foreseen in Ref.\ \cite{Borasoy:1999md}. 

In view of the poor convergence of the chiral series, we do not expect that our approach is the last word on the non-leptonic hyperon decays in the framework of an effective field theory. Nonetheless, it is a clear success that we managed to describe simultaneously the {\it s}- and the {\it p}-wave amplitudes. We regard all our conceptual building blocks as significant for future developments: a modern relativistic framework (EOMS) for loop calculations that respect the power counting; the inclusion of states that lie close in mass to the multiplet of the decaying hyperons; and, finally, the use of up-to-date high-precision data on isospin-decomposed decay amplitudes and strong final-state phase shifts. 

Concerning additional degrees of freedom, we were conservative by including spin-3/2 decuplet states explicitly in the loops, but using other low-lying resonances (of spin 1/2) only at tree level with the purpose of estimating the counterterms. In view of our results that point to a large importance of the spin-1/2 resonances in the numerical results, one can speculate if one should assign to them a corresponding larger importance in the power counting. In other words, maybe one should consider also loop diagrams with such resonances. In this context, it is important to stress the subtle difference of mass scales for pure strong-interaction processes as compared to the case of hyperon decays. The Roper state is about $500\,$MeV away from the nucleon, a relatively large mass difference in pion-nucleon scattering. But the Roper state is only about $250\,$MeV away from the $\Sigma$ baryon. In a strong-interaction process this smaller mass difference is not explicitly probed due to strangeness conservation. But in a weak decay, this mass difference is probed explicitly and is comparable to the mass splitting within the ground-state multiplet. 

On the other hand, including resonances beyond the octet and decuplet ground states as dynamical degrees of freedom is not without ambiguities. Our approach, following \cite{Borasoy:1999md,Borasoy:1999nt}, is inspired by the quark-model classification of baryon resonances. But at least for the negative-parity resonances it is also conceivable to implement them through unitarized meson-baryon scattering \cite{Siegel:1988rq,Kaiser:1995cy,Oller:2000ma,Garcia-Recio:2003ejq,Kolomeitsev:2003kt,Magas:2005vu,Jido:2007sm,Bruns:2010sv,Mai:2012dt,Doring:2025sgb}. This can easily lead to a double-counting problem. An important driving term for the dynamical creation of hadron states is the Weinberg-Tomozawa interaction \cite{Weinberg:1966kf,Tomozawa:1966jm}. In our diagrams of Figs.\ \ref{fig:s-wave-diags} and \ref{fig:p-wave-diags}, this gives rise to the strong meson-baryon four-point vertices (the dots denote strong, the boxes weak vertices). If the resonance in Fig.\ \ref{1d} was dynamically generated by a series of meson-baryon bubble diagrams, then the diagrams \ref{1b} and \ref{2b} would be the first of such a series. Care will be required to devise a reasonable framework that unites a convincing power counting scheme with a proper handling of the nature of the various relevant degrees of freedom.

It remains to be seen if the influence of the resonance states, the $1/2^-$ and the (excited) $1/2^+$ octets, appears to be significant in other quantities such as semileptonic decay amplitudes. Certainly, our results give further credit to the suggestion of Ref.\ \cite{Borasoy:1999nt} that such states are of crucial importance to understand the weak radiative decays of hyperons. 

In conclusion, our results display the significance of the discussed resonant states in the effective field theory. Although the proliferation of unknown parameters constitutes a significant challenge, the assessment of these states in greater detail has been shown to be important.

\begin{acknowledgements}
The authors acknowledge stimulating discussions with Luis Alvarez-Ruso and Stefan Scherer.
This work was supported in part by the Polish National Science Centre through the Grant No. \\
2019/35/O/ST2/02907, 2023/50/A/ST2/00224, 2024/53/B/ST2/00975, and under the NAWA Foreign Doctoral Fellowship  contract no. PPN/STA/2021/1/00011/U/00001. It has been partially supported by the Spanish Ministerio de Ciencia e Innovación under contracts FIS2017-84038-C2-1-P and PID2020-112777GB-I00 and by Generalitat Valenciana’s CIDE-
GENT/2019/015.
\end{acknowledgements}

\appendix
\appendix

\section{Isospin decompositions}
\label{app:isospin-etc}

By changing a strange quark to a down quark, $s \to d$, or to three quarks, $s \to W^- u \to \bar u d u$, the weak interaction violates isospin. Formally this can be accounted for by the options that the weak process contributes with $\Delta I=1/2$ or $\Delta I = 3/2$. Phenomenologically, the $\Delta I=3/2$ part is suppressed, but not always negligible \cite{AbdEl-Hady:1998qww,Tandean:1998ch,BESIII:2023ldd}. 

In the main part of this paper, we restrict ourselves to $\Delta I = 1/2$ transitions following \cite{Bijnens:1985kj,Jenkins:1991bt}. 
But we present here the general decomposition of the non-leptonic decay amplitudes. One reason for writing this down explicitly is related to the various isospin conventions that are used in the literature. In particular, as already noticed in \cite{Gasser:1984gg}, the decomposition \eqref{eq:meson-field-matrix} does not exactly fit to the mathematical standard conventions. The latter are used, e.g., in the tables provided by the Particle Data Group \cite{ParticleDataGroup:2024cfk}. In the present work we stick to the Clebsch-Gordan coefficients obtained from the use of the particle representations specified in \eqref{eq:3Bgs}, \eqref{eq:decuplet-states}, and \eqref{eq:meson-field-matrix} \cite{Mommers:2022dgw}. 

The four-quark structure $q^a q^c \bar q_b \bar q_d$ of the effective weak interaction \eqref{eq:four-quark-weak} gives rise to a flavor tensor structure $t^{ac}_{bd}$. This
can be decomposed into irreducible representations of $SU_L(3)$: a flavor 27-plet $h^{ac}_{bd}$, a flavor octet $h^a_b$, and a flavor singlet (for group theory details see \cite{Holmberg:2018dtv} and references therein). Being interested in a weak flavor changing process, the singlet is irrelevant. Note that the 27-plet and the octet are traceless. The 27-plet is symmetric in its upper indices and symmetric in its lower indices. For the transition from an $s$ to a $d$ quark, we need a non-vanishing element for $a=2$, $b=3$. The other two quarks of the four-quark operator have the same flavor, leading to a non-vanishing element for $c=d$ for the 27-plet. The condition of being traceless gives us
\begin{eqnarray}
  \label{eq:h27-cond}
  \sum\limits_{c} h^{2c}_{3c} = 0  \,.
\end{eqnarray}
If one is interested in an isospin $\Delta I=1/2$ transition caused by the 27-plet, one needs $h^{21}_{31} = h^{22}_{32}$, because the first upper index ($a=2$) provides already $\Delta I=1/2$. For a $\Delta I=3/2$ transition, on the other hand, one demands the orthogonal relation $h^{21}_{31} = -h^{22}_{32}$. The condition \eqref{eq:h27-cond} fixes then all non-vanishing values (up to an overall normalization).

For the octet, these considerations imply that all matrix elements vanish except for $h^2_3=1$. The process mediated by this choice for $h^a_b$ changes the isospin by $\Delta I=1/2$ (and the strangeness by one unit). For the 27-plet, the choice $h^{21}_{31}=h^{12}_{31}=h^{21}_{13}=h^{12}_{13}=h^{22}_{32}=h^{22}_{23}=1$, $h^{23}_{33}=h^{32}_{33}=-2$ (and all other elements being zero) leads also to $\Delta I=1/2$. The more interesting case is the one that provides a new aspect beyond the octet. The choice \cite{AbdEl-Hady:1998qww}
\begin{eqnarray}
  \label{eq:27-values}
h^{21}_{31}=h^{12}_{31}=h^{21}_{13}=h^{12}_{13}=1/2 \qquad \mbox{and} \qquad  h^{22}_{32}=h^{22}_{23}=-1/2 
\end{eqnarray}
(and all other elements being zero) leads to an isospin change of $\Delta I=3/2$. Whenever we use $h^{ac}_{bd}$ in the following, we have in mind to describe $\Delta I=3/2$ transitions and therefore use (\ref{eq:27-values}). 

Pertinent Clebsch-Gordan coefficients can be determined as follows. We imagine a formal scattering reaction via an intermediate state with fixed isospin $I$. The ``initial state'' is the decaying hyperon and the weak Hamiltonian, represented by the source term $h$. The final state is, of course, the meson-baryon pair. The baryon fields (\ref{eq:3Bgs}) and (\ref{eq:decuplet-states}) can serve as proxies for the respective intermediate state. We use $B$ for total isospin $I=1/2$ and $T$ for $I=3/2$. The Clebsch-Gordan coefficients for the formation of the intermediate state can be obtained from $(B^\dagger)^a_b \{h,B\}^b_a$ for $\Delta I = I =1/2$. On purpose, we wrote $B^\dagger$, because this field is only a flavor proxy for the fictitious intermediate state. It is not supposed to have a spinor structure. For $\Delta I=3/2$, $I=1/2$ we utilize $(B^\dagger)^b_a h^{ac}_{bd} B^d_c$. If we are interested in an intermediate state with $I=3/2$ we take $T^\dagger_{abc} h^b_d B^c_e \epsilon^{ade}$ for the case $\Delta I=1/2$ and $T^\dagger_{abc} h^{ab}_{de} B^e_f \epsilon^{cdf}$ for $\Delta I=3/2$. The formal ``decay'' of the fictitious intermediate state is given by $(\bar B)^a_b \{\Phi,B\}^b_a$ or $(\bar B)^d_a \Phi^e_b T^{abc} \epsilon_{cde}$, respectively. 

As already noted, we use isospin $I$ to classify the two-body final state and $\Delta I$ to classify the isospin change caused by the weak interaction \cite{Overseth:1969bxc}. Thus, we couple formally the isospin of the initial hyperon with $\Delta I$ of the weak Hamiltonian to obtain the final isospin. We denote the partial wave (s or p wave) by $L$ and assign the indices $2\Delta I,2I$. In addition, we split off the phase of the strong final-state interaction $\delta^L_{2I}$.

The decays $\Lambda \to \pi N$ have clearly $I = \Delta I \in \{1/2,3/2\}$. We can write
\begin{eqnarray}
L(\Lambda \to \pi^- p) &=& L^\Lambda_{1,1} e^{i\delta^L_{1}} 
+ \frac{1}{\sqrt{2}} L^\Lambda_{3,3} e^{i\delta^L_{3}} 
\,, \nonumber \\ 
L(\Lambda \to \pi^0 n) &=& 
-\frac{1}{\sqrt{2}} L^\Lambda_{1,1} e^{i\delta^L_{1}}
+ L^\Lambda_{3,3} e^{i\delta^L_{3}}  \,.
    \label{eq:isospin-decomp-Lambda}
\end{eqnarray}
Of course, $\delta^L_{2I}$ refers to the pion-nucleon scattering phase shifts, which are well known \cite{Hoferichter:2015hva}.

The decays $\Xi \to \pi \Lambda$ have always $I=1$. We use the decomposition
\begin{eqnarray}
L(\Xi^- \to \pi^- \Lambda) &=& \left( L^\Xi_{1,1} 
+ \frac{1}{\sqrt{2}} L^\Xi_{3,1} \right) e^{i\delta^L_{1}} 
\,, \nonumber \\ 
L(\Xi^0 \to \pi^0 \Lambda) &=& 
\left(-\frac{1}{\sqrt{2}} L^\Xi_{1,1} 
+ L^\Xi_{3,1} \right) e^{i\delta^L_{1}} .
    \label{eq:isospin-decomp-Xi}
\end{eqnarray}
Here $\delta^L_1$ refers to the pion-$\Lambda$ scattering phase shifts. Unfortunately, those are not so well known \cite{Lu:1994ex,Tandean:2000dx,Meissner:2000re}. 
In principle, there is a second method to access the phase-shift information using the $\Xi$ decays themselves. Since the final state can have only isospin $I=1$, the decay parameter $\phi$  defined in \eqref{eq:phi} is given as
\begin{equation}
\tan\phi=\frac{2sp}{s^2-p^2}\sin(\delta_1^P-\delta_1^S ) \ , \label{eq:phiXi}
\end{equation}
where $s$ and $p$ are real amplitudes obtained by applying the kinematical factors \eqref{eq:sp-def} to the terms given in the parentheses of \eqref{eq:isospin-decomp-Xi}. If the phase shifts are small, then $\phi$ will be proportional to $\delta_1^P-\delta_1^S$ and the proportionality factor is ${2sp}/{(s^2-p^2)}\approx \pm\alpha/\sqrt{1-\alpha^2}$.
In the present work we neglect the phase shifts when extracting the partial wave amplitudes from the data because the current experimental values of $\phi$ are compatible with zero. 

The decays $\Sigma \to \pi N$ have the largest variety of isospin combinations. The initial isospin can couple with $\Delta I=1/2$ and with $\Delta I=3/2$ to a complete isospin that can also be reached by the final two-body state. Thus we have to deal with four amplitudes $L_{2 \Delta I,2I}$ with $\Delta I, I \in \{1/2,3/2\}$. The general decomposition is given by 
\begin{align}
    L{(\Sigma^+\to p\pi^0)}&=
    \frac{\sqrt{2}}{3}\left(  L^\Sigma_{1,1} + \frac12 L^\Sigma_{3,1} \right) \exp\!{(i\delta_{1}^L)} + 
    \left( -
    \frac{\sqrt{2}}{3} L^\Sigma_{1,3} 
    + \frac{2\sqrt{2}}{3} L^\Sigma_{3,3} \right) \exp\!{(i\delta_{3}^L)} \, ,  \nonumber\\
    L{(\Sigma^-\to n\pi^-)}&=
    \left( L^\Sigma_{1,3} + L^\Sigma_{3,3} \right) \exp\!{(i\delta_{3}^L)} \, , \label{eq:Sisospin2-full} \\
    L{(\Sigma^+\to n\pi^+)}&=
     \frac23\left( L^\Sigma_{1,1} + \frac12 L^\Sigma_{3,1} \right) \exp\!{(i\delta_{1}^L)} +
    \left( \frac13 L^\Sigma_{1,3} -\frac23 L^\Sigma_{3,3} \right) \exp\!{(i\delta_{3}^L)} \,,  \nonumber
\end{align}
where we have only included those decay channels that can be measured.\footnote{The $\Sigma^0$ has an electromagnetic decay to $\gamma \Lambda$. Therefore the weak decay branches have so tiny branching ratios that they cannot be observed.}
In the decomposition \eqref{eq:Sisospin2-full}, we have neglected a $\Delta I = 5/2$ piece \cite{Overseth:1969bxc} that cannot be caused in the standard model by the four-quark operators of \eqref{eq:four-quark-weak}.

Unfortunately, there are not enough weak decays to determine all amplitudes from data. In particular, the amplitude $L^\Sigma_{3,1}$ appears always in the same combination with $L^\Sigma_{1,1}$. In total, the relations \eqref{eq:Sisospin2-full} allow to determine the combinations $2 L^\Sigma_{1,1}+L^\Sigma_{3,1}$ and $L^\Sigma_{1,3} -2 L^\Sigma_{3,3}$ from the $\Sigma^+$ decays. Thus we can determine $L^\Sigma_{1,3}$, $L^\Sigma_{3,3}$, and the combination $L^\Sigma_{1,1}+L^\Sigma_{3,1}/2$. Having no way to determine $L^\Sigma_{3,1}$ separately, we neglect it in the present work. At least we can use the results for $L^\Sigma_{3,3}$ to judge the quality of neglecting $\Delta I =3/2$ amplitudes. The same logic has already been applied in \cite{Overseth:1969bxc}.

\section{Integrating out a baryon resonance field}
\label{sec:res-sat}

Consider the generic Lagrangian
\begin{eqnarray}
  \label{eq:gen-res-B}
  {\cal L} &=& \bar R \left( i \slashed{\partial} - m_R \right) R + \bar R j B + \bar B \bar j R + \bar B \left( i \slashed{\partial} - m_B \right) B \,.
\end{eqnarray}
Here, $R$ denotes the resonance field that we want to integrate out. $B$ denotes the baryon field that we want to keep.
The quantity $j$ contains external sources and Goldstone-boson fields. We imagine that $j$ can be split up into a weak part, $j_w$, that is
very small, and a strong part, $j_s$, that is of order ${\cal O}(q)$. Here $q$ denotes the typical small momentum that defines the power counting.
Since $j$ might also contain Gamma matrices, we have introduced the Dirac conjugate $\bar j := \gamma^0 j^\dagger \gamma^0$.
Though we present the discussion for a resonance field of spin 1/2, the generalization to other fermions is straightforward. 

It is important to note
that the mass $m_B$ of $B$ introduces a large scale. The mass $m_R$ of the resonance is supposed to be even larger than $m_B$. From the
point of view of power counting in the effective field theory (EFT), it is important to define how large the mass difference $m_R-m_B$ is
supposed to be. If this mass difference was comparable to the small scales of the EFT, e.g.\ to $q$, then it would be conceivable
that one might not want to
integrate out the resonance field at all, but treat it in the same way as the baryon field $B$. On the other hand, if the coupling of the resonance to
all low-energy degrees of freedom is small and/or if the resonance mass is very high, then it will make sense to integrate out the resonance.
In the following, we will see in particular, how the mass difference $m_R-m_B$ comes into the game. In practice, we will not integrate out the
$\Delta$ decuplet states because their mass differences to the ground-state baryon octet states are small and the coupling constants are large.
We will integrate out, however, the octet resonances from the $N(1440)$ and $N(1535)$ multiplets. One purpose of the present work is to
check how far one comes with this approach that treats those resonances as heavy enough to be integrated out and as important enough to
dominate the low-energy constants. 

We will integrate out the resonance field in the path-integral formalism, but before presenting this line of reasoning it is illustrative to consider the
equation of motion
\begin{eqnarray}
  \label{eq:eom-res}
  \left( i \slashed{\partial} - m_R \right) R = - j B  \,.
\end{eqnarray}
The formal solution is given by 
\begin{eqnarray}
  \label{eq:sol-eom-res}
  R = - \frac1{i \slashed{\partial} - m_R} \left(j B \right)  \,.
\end{eqnarray}
The non-trivial aspect is how to deal with the derivative operator in the denominator. 
In purely mesonic chiral perturbation theory, all derivative operators would produce small quantities of order $q$. Then one can power expand
in the number of derivatives and eventually truncate this series. This, however, is different
in baryonic chiral perturbation theory where a derivative acting on a baryon field can produce a baryon mass, i.e.\ a large scale. On a formal level
it is still possible to power expand in the number of derivatives. But one might not truncate the series right away. For the case at hand, the
derivatives emerging from the derivative expansion of the operator in (\ref{eq:sol-eom-res}) act on the ``mesonic'' quantity $j$ and on the
baryon field $B$. For the latter case one obtains the baryon mass $m_B$. Therefore one finds 
\begin{eqnarray}
  \label{eq:sol-eom-res-exp}
  R = \frac1{m_R - m_B} \, j B + {\cal O}(q j)  \,.
\end{eqnarray}
Here we have used the free equation of motion for the baryon field $B$. 
If one included further interactions of the baryon field $B$ with the low-energy degrees of freedom, the effect could still be expressed
as ${\cal O}(q j)$. Thus we see how the mass difference of the baryons appears in a generic way. Expanding the expression further in powers
of $m_B/m_R$ would only be meaningful if the resonance mass was much larger than the baryon mass.
This is not the case for the resonances that we
consider in this work. We observe in (\ref{eq:sol-eom-res-exp}) an interesting competition of importance. The smaller the interaction strength
encoded in $j$, the less important the whole contribution is. The smaller the mass difference $m_R - m_B$, the more important the contribution
gets.

The final step is to see the consequences of the resonance for the low-energy dynamics. To this end, we determine the path integral over the
resonance fields,
\begin{eqnarray}
  \label{eq:pathint-R}
  \int {\cal D}\bar R \, {\cal D}R \, \exp\left[ i \int d^4 x {\cal L} \right]
\end{eqnarray}
where the Lagrangian is given in (\ref{eq:gen-res-B}). It is of advantage to introduce some short-hand notations:
\begin{eqnarray}
  J := j B \,, \qquad \bar R \otimes A \otimes R := \int d^4 x \bar R \left( i \slashed{\partial} - m_R \right) R  \,.
  \label{eq:shorthands-path-int}  
\end{eqnarray}
This allows us to write
\begin{eqnarray}
  && \int {\cal D}\bar R \, {\cal D}R \, \exp\left[ i \bar R \otimes A \otimes R + i \bar R \otimes J + i \bar J \otimes R \right] \nonumber \\
  &=& \int {\cal D}\bar R \, {\cal D}R \, \exp\left[ i \left(\bar R + \bar J \otimes A^{-1} \right) \otimes A \otimes \left(R + A^{-1} \otimes J \right)
      - i \bar J \otimes A^{-1} \otimes J \right]  \nonumber \\
  & \sim & \exp\left[ - i \bar J \otimes A^{-1} \otimes J \right]  \,.
  \label{eq:completing-square}  
\end{eqnarray}
Translating back to the explicit four-dimensional effective action, we obtain the following contribution to the corresponding effective Lagrangian:
\begin{eqnarray}
  - \bar B \bar j \frac1{i \slashed{\partial} - m_R} \left(j B \right)
  = \frac1{m_R-m_B} \, \bar B \bar j \, j B + {\cal O}(q \, j^2) \,.
  \label{eq:eff-lagr-fromR}  
\end{eqnarray}
We see that the coefficient $1/(m_R-m_B)$ can be interpreted as a low-energy constant for the operator structure $\bar B \bar j \, j B$. 

As a last step, we decompose $j=j_w+j_s$ and discuss the various terms that emerge. Integrating out a baryon resonance in the sector of the
strong interaction yields
\begin{eqnarray}
  \label{eq:only-strong}
  \frac1{m_R-m_B} \, \bar B \bar j_s \, j_s B + {\cal O}(q^3) \,.
\end{eqnarray}
The explicit term is of order ${\cal O}(q^2)$ and describes at least a four-point interaction. This is the traditional use of resonance saturation
for the ${\cal O}(q^2)$ low-energy constants.

We obtain also terms of quadratic order in the weak interaction. We are not interested in these very tiny interaction terms. Finally, we obtain
terms that are linear in the weak interaction: 
\begin{eqnarray}
  \label{eq:bilinear-ws}
  \frac1{m_R-m_B} \, \bar B (\bar j_s \, j_w + \bar j_w \, j_s) B + {\cal O}(q^2) \,.
\end{eqnarray}
The explicit term is of order ${\cal O}(q)$ and describes at least a three-point interaction. These are exactly the terms that matter for our
applications where a baryon decays into another baryon and a pion. Here the momentum $q$ refers to the momentum of the pion in the
rest frame of any of the two baryons. In powers of kaon masses, this momentum is of order $M_K^2$. Integrating out the spin-1/2 resonances
yields tree-level terms (\ref{eq:bilinear-ws}). In the spirit of resonance saturation, the corresponding coefficients are supposed to provide
reasonable estimates for the otherwise unknown low-energy constants of the weak interaction Lagrangian at order ${\cal O}(q)$.

Finally, we note that at the level of accuracy that we pursue, the following three methods are completely equivalent:
1.\ Integrating out a resonance field along the lines of (\ref{eq:completing-square}), (\ref{eq:eff-lagr-fromR}).
2.\ Using the solution (\ref{eq:sol-eom-res}) of the equation of motion and plugging it back into the original Lagrangian (\ref{eq:gen-res-B}).
3.\ Using the original Lagrangian right away to generate tree-level diagrams with intermediate resonance propagators.

\section{Determining strong coupling constants from partial decay widths}
\label{app:det-strong-Roper}
We consider the following $J^P = 1/2^+$ states: the ground-state octet (\ref{eq:3Bgs})
and the Roper octet
\begin{equation}
  R_{+} = \left(
    \begin{array}{ccc}
      \frac{1}{\sqrt{2}} \Sigma^0_+ + \frac{1}{\sqrt{6}} \Lambda_+ & \Sigma^+_+ & N^+_+ \\
      \Sigma^-_+ & -\frac{1}{\sqrt{2}} \Sigma^0_+ + \frac{1}{\sqrt{6}} \Lambda_+ & N^0_+  \\
      \Xi^-_+ & \Xi^0_+ & -\frac{2}{\sqrt{6}} \Lambda_+ 
    \end{array}
  \right) \,.
    \label{eq:3res}
\end{equation}

We want to describe the strong decays of a state from the Roper multiplet to a baryon ground state and a Goldstone boson. The corresponding three-point couplings between baryons and Goldstone bosons are provided at LO (first order in meson-field derivatives) by the
Lagrangian (\ref{eq:L1phiB*B}).
For our purposes we have $u_\mu \to 2 \partial_\mu \phi/F_{\pi,K,\eta}$ with the meson matrix (\ref{eq:meson-field-matrix}).
For the fits, we use $F_\pi=92.1 \,$MeV and $F_K = 110\,$MeV \cite{ParticleDataGroup:2024cfk}. Note that the meson decay constants have relatively
large flavor breaking corrections. Therefore we use $F_\pi$ only for the pions but $F_K$ for the kaons. 

The coupling constants $D^*$ and $F^*$ are determined (in LO) from the decays $R_{+} \to B \phi$. 
The generic Lagrangian for the decay is (here the fields do not denote the matrices but rather specific states) 
\begin{eqnarray}
  \label{eq:decay-lagr-gen}
  {\cal L} = -c_{R_{+}B\phi} \bar B \gamma^\mu \gamma_5 R_{+} \partial_\mu \phi
  = c_{R_{+}B\phi} \, \partial_\mu \left( \bar B \gamma^\mu \gamma_5 R_{+} \right) \phi
  = c_{R_{+}B\phi} \, (m_B+m_{R_{+}}) \bar B i \gamma_5 R_{+} \phi \,.
\end{eqnarray}
This leads to the spin averaged squared matrix element
\begin{equation}
  \label{eq:matr2av}
  \langle \vert {\cal M} \vert^2 \rangle = c_{R_{+}B\phi}^2 \, (m_B+m_{R_{+}})^2 \, 2 m_{R_{+}} \, (E_B-m_B)
  = c_{R_{+}B\phi}^2 \, (m_B+m_{R_{+}})^2 \left( (m_{R_{+}}-m_B)^2 - M_\phi^2 \right)  \,. 
\end{equation}
Here $E_B$ denotes the energy of the outgoing baryon in the rest frame of the decaying resonance $R_{+}$. In the non-relativistic limit $E_B-m_B$
provides the p-wave factor of a squared three-momentum. 
The decay width is given by
\begin{eqnarray}
  \Gamma_{R_{+}B\phi} = c_{R_{+}B\phi}^2 \, (m_B+m_{R_{+}})^2 \, \frac{(m_{R_{+}}-m_B)^2 -M_\phi^2}{8 \pi \, m_{R_{+}}^2} \, \frac{\lambda^{1/2}(m_{R_{+}}^2,m_B^2,M_\phi^2)}{2 m_{R_{+}}}  
  \label{eq:width-strong-pwave3}
\end{eqnarray}
with the K\"all\'en function
\begin{eqnarray}
  \label{eq:kallenfunc}
  \lambda(a,b,c):=a^2+b^2+c^2-2(ab+bc+ac) \,.
\end{eqnarray}

Next we turn to the flavor factors relevant for the measured decays. We find     
\begin{eqnarray}
  c_{N_+ N \pi} &=& \frac3{8 F_\pi^2} (D^*+F^*)    \,,  \nonumber \\ 
  c_{\Lambda_+ N K}  &=& \frac1{12 F_K^2} (D^*+3 F^*)   \,,  \nonumber \\ 
  c_{\Lambda_+ \Sigma \pi} &=& \frac1{2 F_\pi^2} D^*    \,,  \nonumber \\ 
  c_{\Sigma_+ N K} &=& \frac{1}{4 F_K^2} (D^*-F^*)   \,,  \nonumber \\ 
  c_{\Sigma_+ \Lambda \pi} &=& \frac1{6 F_\pi^2} D^*    \,,  \nonumber \\ 
  c_{\Sigma_+ \Sigma \pi} &=& \frac1{F_\pi^2} F^*     \,.  
  \label{eq:flavor-fac}
\end{eqnarray}

As input to determine the coupling constants, we use the Breit-Wigner masses and widths and the branching ratios as determined from the 
Bonn-Gatchina partial-wave analysis \cite{CBELSATAPS:2015kka,Sarantsev:2019xxm}.
We identify $\Lambda(1600)$ and $\Sigma(1660)$ as members of the Roper multiplet.
The situation is rather unclear for the cascade state. 
We use the Gell-Mann--Okubo relation (\ref{eq:gell-okubo}) to estimate the mass of the cascade state $\Xi_+$. We find $m_{\Xi_+} \approx 1.81\,$GeV. 

A fit to the six measured decays yields
\begin{eqnarray}
  \label{eq:resDRFR}
  D^* \approx 0.713 \,, \qquad F^* \approx 0.342 \,.
\end{eqnarray}
This should be compared to the results used by Borasoy and Holstein \cite{Borasoy:1999md} given as $D^* = 0.60$ and $F^*=0.11$ based on older decay data \cite{Borasoy:1996bx}. In particular, the value for $F^*$ is very different. 

With the values \eqref{eq:resDRFR}, we find a very good agreement with the branching ratios collected in Table \ref{table:radii-res} (using the total widths from experiment).
\begin{table}[h]
	\centering	
	\begin{tabular}{|l||c|c|}
          \hline 
          & expt. & calc.   \\
          \hline 
          \hline 
          $N_+ \to N \pi$ & $0.63 \pm 0.02$ &  0.64  \\
          \hline 
          $\Lambda_+ \to N K$ & $0.29 \pm 0.06$ &  0.29 \\
          \hline 
          $\Lambda_+ \to \Sigma \pi$ & $0.37 \pm 0.07$ & 0.38 \\
          \hline 
          $\Sigma_+ \to N K$ & $0.07\pm 0.03$ & 0.05 \\
          \hline 
          $\Sigma_+ \to \Lambda \pi$ & $0.35 \pm 0.12$ & 0.23 \\ 
          \hline 
          $\Sigma_+ \to \Sigma \pi$ & $0.37 \pm 0.10$ & 0.21 \\
          \hline 
      \end{tabular}
	\caption{Comparison of calculated (calc.) and measured (expt.) \cite{CBELSATAPS:2015kka,Sarantsev:2019xxm} branching ratios. }
	\label{table:radii-res}
\end{table}

\bibliographystyle{apsrev4-2}
\bibliography{references}

@article{Weinberg:1966kf,
    author = "Weinberg, Steven",
    title = "{Pion scattering lengths}",
    doi = "10.1103/PhysRevLett.17.616",
    journal = "Phys. Rev. Lett.",
    volume = "17",
    pages = "616--621",
    year = "1966"
}

@article{Tomozawa:1966jm,
    author = "Tomozawa, Y.",
    title = "{Axial vector coupling renormalization and the meson baryon scattering lengths}",
    doi = "10.1007/BF02857517",
    journal = "Nuovo Cim. A",
    volume = "46",
    pages = "707--717",
    year = "1966"
}

@article{Bertilsson:2023htb,
    author = "Bertilsson, Magnus and Leupold, Stefan",
    title = "{Goldberger-Treiman relation and Wu-type experiment in the decuplet sector}",
    eprint = "2311.07471",
    archivePrefix = "arXiv",
    primaryClass = "hep-ph",
    doi = "10.1103/PhysRevD.109.034028",
    journal = "Phys. Rev. D",
    volume = "109",
    number = "3",
    pages = "034028",
    year = "2024"
}

@misc{Kummer:2025kch,
    author = "Kummer, Christian and Leupold, Stefan and von Smekal, Lorenz",
    title = "{Kinetic Mixing and Axial Charges in the Parity-Doublet Model}",
    eprint = "2512.03894",
    archivePrefix = "arXiv",
    primaryClass = "hep-ph",
    ANmonth = "12",
    year = "2025"
}

@article{Jido:2001nt,
    author = "Jido, Daisuke and Oka, Makoto and Hosaka, Atsushi",
    title = "{Chiral symmetry of baryons}",
    eprint = "hep-ph/0110005",
    archivePrefix = "arXiv",
    doi = "10.1143/PTP.106.873",
    journal = "Prog. Theor. Phys.",
    volume = "106",
    pages = "873--908",
    year = "2001"
}

@article{Detar:1988kn,
    author = "Detar, Carleton E. and Kunihiro, Teiji",
    title = "{Linear $\sigma$ Model With Parity Doubling}",
    reportNumber = "RIFP-748",
    doi = "10.1103/PhysRevD.39.2805",
    journal = "Phys. Rev. D",
    volume = "39",
    pages = "2805",
    year = "1989"
}

@article{Bruns:2010sv,
    author = "Bruns, Peter C. and Mai, Maxim and Mei{\ss}ner, Ulf-G.",
    title = "{Chiral dynamics of the S11(1535) and S11(1650) resonances revisited}",
    eprint = "1012.2233",
    archivePrefix = "arXiv",
    primaryClass = "nucl-th",
    reportNumber = "HISKP-TH-10-28, FZJ-IKP-TH-2010-25",
    doi = "10.1016/j.physletb.2011.02.008",
    journal = "Phys. Lett. B",
    volume = "697",
    pages = "254--259",
    year = "2011"
}

@article{Doring:2025sgb,
    author = {D{\"o}ring, Michael and Haidenbauer, Johann and Mai, Maxim and Sato, Toru},
    title = "{Dynamical coupled-channel models for hadron dynamics}",
    eprint = "2505.02745",
    archivePrefix = "arXiv",
    primaryClass = "nucl-th",
    reportNumber = "JLAB-THY-25-4298",
    doi = "10.1016/j.ppnp.2025.104213",
    journal = "Prog. Part. Nucl. Phys.",
    volume = "146",
    pages = "104213",
    year = "2026"
}

@article{Jido:2007sm,
    author = "Jido, D. and D{\"o}ring, M. and Oset, E.",
    title = "{Transition form factors of the N*(1535) as a dynamically generated resonance}",
    eprint = "0712.0038",
    archivePrefix = "arXiv",
    primaryClass = "nucl-th",
    reportNumber = "YITP-07-83",
    doi = "10.1103/PhysRevC.77.065207",
    journal = "Phys. Rev. C",
    volume = "77",
    pages = "065207",
    year = "2008"
}

@article{Oller:2000ma,
    author = "Oller, J. A. and Oset, E. and Ramos, A.",
    title = "{Chiral unitary approach to meson-meson and meson-baryon interactions and nuclear applications}",
    eprint = "hep-ph/0002193",
    archivePrefix = "arXiv",
    reportNumber = "FZJ-IKP-TH-1999-37, FTUV-99-1215, IFIC-99-1215",
    doi = "10.1016/S0146-6410(00)00104-6",
    journal = "Prog. Part. Nucl. Phys.",
    volume = "45",
    pages = "157--242",
    year = "2000"
}

@article{Kaiser:1995cy,
    author = "Kaiser, Norbert and Siegel, P. B. and Weise, W.",
    title = "{Chiral dynamics and the S11(1535) nucleon resonance}",
    eprint = "nucl-th/9507036",
    archivePrefix = "arXiv",
    reportNumber = "TUM-T39-95-10",
    doi = "10.1016/0370-2693(95)01203-3",
    journal = "Phys. Lett. B",
    volume = "362",
    pages = "23--28",
    year = "1995"
}

@article{Gasser:1983yg,
    author = "Gasser, J. and Leutwyler, H.",
    title = "{Chiral Perturbation Theory to One Loop}",
    reportNumber = "CERN-TH-3689",
    doi = "10.1016/0003-4916(84)90242-2",
    journal = "Annals Phys.",
    volume = "158",
    pages = "142",
    year = "1984"
}

@article{Lu:1994ex,
    author = "Lu, Ming and Wise, Mark B. and Savage, Martin J.",
    title = "{Strong $\Lambda \pi$ phase shifts for CP violation in weak $\Xi \to \Lambda \pi$ decay}",
    eprint = "hep-ph/9407260",
    archivePrefix = "arXiv",
    reportNumber = "CALT-68-1940, CMU-HEP-94-22",
    doi = "10.1016/0370-2693(94)91456-7",
    journal = "Phys. Lett. B",
    volume = "337",
    pages = "133--136",
    year = "1994"
}

@article{Tandean:2000dx,
    author = "Tandean, Jusak and Thomas, Anthony William and Valencia, German E",
    title = "{Can the Lambda pi scattering phase shifts be large?}",
    eprint = "hep-ph/0011214",
    archivePrefix = "arXiv",
    reportNumber = "UK-TP-00-06, ISU-HET-00-6, ADP-00-52-T432",
    doi = "10.1103/PhysRevD.64.014005",
    journal = "Phys. Rev. D",
    volume = "64",
    pages = "014005",
    year = "2001"
}

@article{Meissner:2000re,
    author = "Mei{\ss}ner, Ulf-G. and Oller, Jose Antonio",
    title = "{The S wave Lambda pi phase shift is not large}",
    eprint = "hep-ph/0011293",
    archivePrefix = "arXiv",
    reportNumber = "FZJ-IKP-TH-2000-29",
    doi = "10.1103/PhysRevD.64.014006",
    journal = "Phys. Rev. D",
    volume = "64",
    pages = "014006",
    year = "2001"
}

@article{Siegel:1988rq,
    author = "Siegel, P. B. and Weise, W.",
    title = "{Low-energy $K^-$ Nucleon Potentials and the Nature of the $\Lambda(1405)$}",
    reportNumber = "TPR-88-4",
    doi = "10.1103/PhysRevC.38.2221",
    journal = "Phys. Rev. C",
    volume = "38",
    pages = "2221--2229",
    year = "1988"
}

@article{Garcia-Recio:2003ejq,
    author = "Garcia-Recio, C. and Lutz, M. F. M. and Nieves, J.",
    title = "{Quark mass dependence of s wave baryon resonances}",
    eprint = "nucl-th/0305100",
    archivePrefix = "arXiv",
    reportNumber = "GSI-PREPRINT-2003-16",
    doi = "10.1016/j.physletb.2003.11.073",
    journal = "Phys. Lett. B",
    volume = "582",
    pages = "49--54",
    year = "2004"
}

@article{Magas:2005vu,
    author = "Magas, V. K. and Oset, E. and Ramos, A.",
    title = "{Evidence for the Two-Pole Structure of the $\Lambda(1405)$ Resonance}",
    eprint = "hep-ph/0503043",
    archivePrefix = "arXiv",
    doi = "10.1103/PhysRevLett.95.052301",
    journal = "Phys. Rev. Lett.",
    volume = "95",
    pages = "052301",
    year = "2005"
}

@article{Mai:2012dt,
    author = "Mai, Maxim and Mei{\ss}ner, Ulf-G.",
    title = "{New insights into antikaon-nucleon scattering and the structure of the Lambda(1405)}",
    eprint = "1202.2030",
    archivePrefix = "arXiv",
    primaryClass = "nucl-th",
    doi = "10.1016/j.nuclphysa.2013.01.032",
    journal = "Nucl. Phys. A",
    volume = "900",
    pages = "51 -- 64",
    year = "2013"
}

@article{Hall:2014uca,
    author = "Hall, Jonathan M. M. and Kamleh, Waseem and Leinweber, Derek B. and Menadue, Benjamin J. and Owen, Benjamin J. and Thomas, Anthony W. and Young, Ross D.",
    title = "{Lattice QCD Evidence that the {\ensuremath{\Lambda}}(1405) Resonance is an Antikaon-Nucleon Molecule}",
    eprint = "1411.3402",
    archivePrefix = "arXiv",
    primaryClass = "hep-lat",
    reportNumber = "ADP-14-34-T893",
    doi = "10.1103/PhysRevLett.114.132002",
    journal = "Phys. Rev. Lett.",
    volume = "114",
    number = "13",
    pages = "132002",
    year = "2015"
}

@article{Beane:2002ud,
    author = "Beane, Silas R. and van Kolck, Ubirajara",
    title = "{The Role of the Roper in QCD}",
    eprint = "nucl-th/0212039",
    archivePrefix = "arXiv",
    reportNumber = "INT-PUB-02-52",
    doi = "10.1088/0954-3899/31/8/021",
    journal = "J. Phys. G",
    volume = "31",
    pages = "921--934",
    year = "2005"
}

@article{Geng:2008mf,
    author = "Geng, L. S. and Martin Camalich, J. and Alvarez-Ruso, L. and Vicente Vacas, M. J.",
    title = "{Leading SU(3)-breaking corrections to the baryon magnetic moments in Chiral Perturbation Theory}",
    eprint = "0805.1419",
    archivePrefix = "arXiv",
    primaryClass = "hep-ph",
    doi = "10.1103/PhysRevLett.101.222002",
    journal = "Phys. Rev. Lett.",
    volume = "101",
    pages = "222002",
    year = "2008"
}

@article{Kolomeitsev:2003kt,
    author = "Kolomeitsev, E. E. and Lutz, M. F. M.",
    title = "{On baryon resonances and chiral symmetry}",
    eprint = "nucl-th/0305101",
    archivePrefix = "arXiv",
    reportNumber = "GSI-PREPRINT-2003-17",
    doi = "10.1016/j.physletb.2004.01.066",
    journal = "Phys. Lett. B",
    volume = "585",
    pages = "243--252",
    year = "2004"
}

@article{Lutz:2001yb,
    author = "Lutz, M. F. M. and Kolomeitsev, E. E.",
    title = "{Relativistic chiral SU(3) symmetry, large N(c) sum rules and meson baryon scattering}",
    eprint = "nucl-th/0105042",
    archivePrefix = "arXiv",
    reportNumber = "GSI-PREPRINT-2001-12, ECT-2001-10",
    doi = "10.1016/S0375-9474(01)01312-4",
    journal = "Nucl. Phys. A",
    volume = "700",
    pages = "193--308",
    year = "2002"
}

@article{Semke:2005sn,
    author = "Semke, A. and Lutz, M. F. M.",
    title = "{Baryon self energies in the chiral loop expansion}",
    eprint = "nucl-th/0511061",
    archivePrefix = "arXiv",
    doi = "10.1016/j.nuclphysa.2006.07.043",
    journal = "Nucl. Phys. A",
    volume = "778",
    pages = "153--180",
    year = "2006"
}

@article{Pascalutsa:2006up,
    author = "Pascalutsa, Vladimir and Vanderhaeghen, Marc and Yang, Shin Nan",
    title = "{Electromagnetic excitation of the Delta(1232)-resonance}",
    eprint = "hep-ph/0609004",
    archivePrefix = "arXiv",
    reportNumber = "JLAB-THY-06-537",
    doi = "10.1016/j.physrep.2006.09.006",
    journal = "Phys. Rept.",
    volume = "437",
    pages = "125--232",
    year = "2007"
}

@article{Crede:2024hur,
    author = "Crede, Volker and Yelton, John",
    title = "{70 years of hyperon spectroscopy: a review of strange $\Xi$, $\Omega$ baryons, and the spectrum of charmed and bottom baryons}",
    eprint = "2502.08815",
    archivePrefix = "arXiv",
    primaryClass = "hep-ex",
    doi = "10.1088/1361-6633/ad7610",
    journal = "Rept. Prog. Phys.",
    volume = "87",
    number = "10",
    pages = "106301",
    year = "2024"
}

@unpublished{Gell-Mann:1961omu,
  author = {Gell-Mann, Murray},
  title = {The Eightfold Way: A theory of strong interaction symmetry},
  note = {{California} Institute of Technology Report CTSL-20, TID-12608},
  doi = "10.2172/4008239",
  year = {1961}
}

@article{Okubo:1961jc,
    author = "Okubo, Susumu",
    title = "{Note on unitary symmetry in strong interactions}",
    doi = "10.1143/PTP.27.949",
    journal = "Prog. Theor. Phys.",
    volume = "27",
    pages = "949--966",
    year = "1962"
}

@article{Okubo:1962zzc,
    author = "Okubo, S.",
    title = "{Note on Unitary Symmetry in Strong Interaction. II Excited States of Baryons}",
    doi = "10.1143/PTP.28.24",
    journal = "Prog. Theor. Phys.",
    volume = "28",
    pages = "24--32",
    year = "1962"
}

@article{Cirigliano:2011ny,
    author = "Cirigliano, Vincenzo and Ecker, Gerhard and Neufeld, Helmut and Pich, Antonio and Portoles, Jorge",
    title = "{Kaon Decays in the Standard Model}",
    eprint = "1107.6001",
    archivePrefix = "arXiv",
    primaryClass = "hep-ph",
    reportNumber = "FTUV-11-0729, IFIC-11-02, UWTHPH-2011-25",
    doi = "10.1103/RevModPhys.84.399",
    journal = "Rev. Mod. Phys.",
    volume = "84",
    pages = "399",
    year = "2012"
}

@article{RBC:2020kdj,
    author = "Abbott, R. and others",
    collaboration = "RBC, UKQCD",
    title = "{Direct CP violation and the $\Delta I=1/2$ rule in $K\to\pi\pi$ decay from the standard model}",
    eprint = "2004.09440",
    archivePrefix = "arXiv",
    primaryClass = "hep-lat",
    reportNumber = "CERN-TH-2020-058, MIT-CTP/5197",
    doi = "10.1103/PhysRevD.102.054509",
    journal = "Phys. Rev. D",
    volume = "102",
    number = "5",
    pages = "054509",
    year = "2020"
}

@article{Ecker:1988te,
      author         = "Ecker, G. and Gasser, J. and Pich, A. and de Rafael, E.",
      title          = "{The Role of Resonances in Chiral Perturbation Theory}",
      journal        = "Nucl. Phys.",
      volume         = "B321",
      year           = "1989",
      pages          = "311-342",
      doi            = "10.1016/0550-3213(89)90346-5",
      reportNumber   = "CERN-TH-5185/88, UWThPh-1988-29, BUTP-88/18,
                        CPT-88/PE-2158",
      SLACcitation   = "%%CITATION = NUPHA,B321,311;%%"
}

@article{Donoghue:1988ed,
    author = "Donoghue, John F. and Ramirez, Carlos and Valencia, German",
    title = "{The Spectrum of QCD and Chiral Lagrangians of the Strong and Weak Interactions}",
    reportNumber = "UMHEP-298",
    doi = "10.1103/PhysRevD.39.1947",
    journal = "Phys. Rev. D",
    volume = "39",
    pages = "1947",
    year = "1989"
}

@article{Kampf:2011ty,
    author = "Kampf, Karol and Novotny, Jiri",
    title = "{Resonance saturation in the odd-intrinsic parity sector of low-energy QCD}",
    eprint = "1104.3137",
    archivePrefix = "arXiv",
    primaryClass = "hep-ph",
    reportNumber = "LU-TP-11-20",
    doi = "10.1103/PhysRevD.84.014036",
    journal = "Phys. Rev. D",
    volume = "84",
    pages = "014036",
    year = "2011"
}

@article{Epelbaum:2001fm,
    author = "Epelbaum, Evgeny and {Mei\ss}ner, Ulf-G. and Gloeckle, Walter and Elster, Charlotte",
    title = "{Resonance saturation for four nucleon operators}",
    eprint = "nucl-th/0106007",
    archivePrefix = "arXiv",
    reportNumber = "FZJ-IKP-TH-2001-09, FZJ-IKP(TH)-2001-09",
    doi = "10.1103/PhysRevC.65.044001",
    journal = "Phys. Rev. C",
    volume = "65",
    pages = "044001",
    year = "2002"
}

@article{Ecker:1989yg,
      author         = "Ecker, G. and Gasser, J. and Leutwyler, H. and Pich, A.
                        and de Rafael, E.",
      title          = "{Chiral Lagrangians for Massive Spin 1 Fields}",
      journal        = "Phys. Lett.",
      volume         = "B223",
      year           = "1989",
      pages          = "425-432",
      doi            = "10.1016/0370-2693(89)91627-4",
      reportNumber   = "BUTP-89/4-BERN, UWThPh-1989-9, FTUV/89-7, CPT-89/PE-2243",
      SLACcitation   = "%%CITATION = PHLTA,B223,425;%%"
}

@Book{SM-Donoghue,
     author    = "Donoghue, John F. and Golowich, Eugene and Holstein, Barry R.",
     title     = "Dynamics of the Standard Model",
     ANaddress   = "Boulder, CO",
     publisher = {Cambridge University Press},
     year      = "2014"
}

@article{Overseth:1969bxc,
    author = "Overseth, O. E. and Pakvasa, S.",
    title = "{Final-state interactions in nonleptonic hyperon decay}",
    doi = "10.1103/PhysRev.184.1663",
    journal = "Phys. Rev.",
    volume = "184",
    pages = "1663--1667",
    year = "1969"
}

@article{Hoferichter:2015hva,
    author = "Hoferichter, Martin and Ruiz de Elvira, Jacobo and Kubis, Bastian and Mei{\ss}ner, Ulf-G.",
    title = "{Roy{\textendash}Steiner-equation analysis of pion{\textendash}nucleon scattering}",
    eprint = "1510.06039",
    archivePrefix = "arXiv",
    primaryClass = "hep-ph",
    reportNumber = "INT-PUB-15-050",
    doi = "10.1016/j.physrep.2016.02.002",
    journal = "Phys. Rept.",
    volume = "625",
    pages = "1--88",
    year = "2016"
}

@article{AbdEl-Hady:1998qww,
    author = "Abd El-Hady, A. and Tandean, Jusak and Valencia, G.",
    title = "{Chiral perturbation theory for |Delta I| = 3/2 hyperon decays}",
    eprint = "hep-ph/9808322",
    archivePrefix = "arXiv",
    reportNumber = "ISU-HET-98-2",
    doi = "10.1016/S0375-9474(99)00128-1",
    journal = "Nucl. Phys. A",
    volume = "651",
    pages = "71--89",
    year = "1999"
}

@article{Tandean:1998ch,
    author = "Tandean, Jusak and Valencia, G.",
    title = "{|Delta I| = 3/2 decays of the $\Omega^-$ in chiral perturbation theory}",
    eprint = "hep-ph/9810201",
    archivePrefix = "arXiv",
    reportNumber = "ISU-HET-98-4",
    doi = "10.1016/S0370-2693(99)00277-4",
    journal = "Phys. Lett. B",
    volume = "452",
    pages = "395--401",
    year = "1999"
}

@article{BESIII:2023ldd,
    author = "Ablikim, Medina and others",
    collaboration = "BESIII",
    title = "{Measurements of the absolute branching fractions of $\Omega^-$ decays and test of the $\Delta I=1/2$ rule}",
    eprint = "2309.06368",
    archivePrefix = "arXiv",
    primaryClass = "hep-ex",
    doi = "10.1103/PhysRevD.108.L091101",
    journal = "Phys. Rev. D",
    volume = "108",
    number = "9",
    pages = "L091101",
    year = "2023"
}

@article{Borasoy:1999nt,
    author = "Borasoy, Bugra and Holstein, Barry R.",
    title = "{Resonances in radiative hyperon decays}",
    eprint = "hep-ph/9902431",
    archivePrefix = "arXiv",
    doi = "10.1103/PhysRevD.59.054019",
    journal = "Phys. Rev. D",
    volume = "59",
    pages = "054019",
    year = "1999"
}

@article{Bijnens:1985kj,
    author = "Bijnens, J. and Sonoda, H. and Wise, Mark B.",
    title = "{On the Validity of Chiral Perturbation Theory for Weak Hyperon Decays}",
    reportNumber = "CALT-68-1221",
    doi = "10.1016/0550-3213(85)90569-3",
    journal = "Nucl. Phys. B",
    volume = "261",
    pages = "185--198",
    year = "1985"
}

@article{ParticleDataGroup:2024cfk,
    author = "Navas, S. and others",
    collaboration = "Particle Data Group",
    title = "{Review of particle physics}",
    doi = "10.1103/PhysRevD.110.030001",
    journal = "Phys. Rev. D",
    volume = "110",
    number = "3",
    pages = "030001",
    year = "2024"
}

@article{Scherer:2012xha,
      author         = "Scherer, Stefan and Schindler, Matthias R.",
      title          = "{A Primer for Chiral Perturbation Theory}",
      journal        = "Lect. Notes Phys.",
      volume         = "830",
      year           = "2012",
      ALTpages          = "pp.1-338",
      doi            = "10.1007/978-3-642-19254-8",
      SLACcitation   = "%%CITATION = LNPHA,830,pp.1;%%"
}

@article{Mommers:2022dgw,
    author = "Mommers, Cornelis J. G. and Leupold, Stefan",
    title = "{Estimates for rare three-body decays of the \ensuremath{\Omega} baryon using chiral symmetry and the \ensuremath{\Delta}I=1/2 rule}",
    eprint = "2208.11078",
    archivePrefix = "arXiv",
    primaryClass = "hep-ph",
    doi = "10.1103/PhysRevD.106.093001",
    journal = "Phys. Rev. D",
    volume = "106",
    number = "9",
    pages = "093001",
    year = "2022"
}

@article{Flores-Mendieta:2019lao,
    author = "Flores-Mendieta, Rub\'en",
    title = "{$s$-wave baryon nonleptonic decay amplitude in large-$N_c$ chiral perturbation theory}",
    eprint = "1902.05602",
    archivePrefix = "arXiv",
    primaryClass = "hep-ph",
    doi = "10.1103/PhysRevD.99.094033",
    journal = "Phys. Rev. D",
    volume = "99",
    number = "9",
    pages = "094033",
    year = "2019"
}

@article{AbdEl-Hady:1999llb,
    author = "Abd El-Hady, A. and Tandean, Jusak",
    title = "{Hyperon nonleptonic decays in chiral perturbation theory reexamined}",
    eprint = "hep-ph/9908498",
    archivePrefix = "arXiv",
    reportNumber = "ISU-HET-99-10",
    doi = "10.1103/PhysRevD.61.114014",
    journal = "Phys. Rev. D",
    volume = "61",
    pages = "114014",
    year = "2000"
}

@article{Holmberg:2018dtv,
    author = "Holmberg, M\r{a}ns and Leupold, Stefan",
    title = "{The relativistic chiral Lagrangian for decuplet and octet baryons at next-to-leading order}",
    eprint = "1802.05168",
    archivePrefix = "arXiv",
    primaryClass = "hep-ph",
    doi = "10.1140/epja/i2018-12533-3",
    journal = "Eur. Phys. J. A",
    volume = "54",
    number = "6",
    pages = "103",
    year = "2018"
}

@article{Borasoy:1998ku,
    author = "Borasoy, B. and Holstein, Barry R.",
    title = "{Nonleptonic hyperon decays in chiral perturbation theory}",
    eprint = "hep-ph/9805430",
    archivePrefix = "arXiv",
    doi = "10.1007/s100520050323",
    journal = "Eur. Phys. J. C",
    volume = "6",
    pages = "85--107",
    year = "1999"
}

@article{Jenkins:1991bt,
    author = "Jenkins, Elizabeth Ellen",
    title = "{Hyperon nonleptonic decays in chiral perturbation theory}",
    reportNumber = "UCSD-PTH-91-24",
    doi = "10.1016/0550-3213(92)90111-N",
    journal = "Nucl. Phys. B",
    volume = "375",
    pages = "561--581",
    year = "1992"
}

@article{Borasoy:1999md,
    author = "Borasoy, Bugra and Holstein, Barry R.",
    title = "{The Role of resonances in nonleptonic hyperon decays}",
    eprint = "hep-ph/9902351",
    archivePrefix = "arXiv",
    doi = "10.1103/PhysRevD.59.094025",
    journal = "Phys. Rev. D",
    volume = "59",
    pages = "094025",
    year = "1999"
}

@article{BESIII:2018cnd,
    author = "Ablikim, M. and others",
    collaboration = "BESIII",
    title = "{Polarization and Entanglement in Baryon-Antibaryon Pair Production in Electron-Positron Annihilation}",
    eprint = "1808.08917",
    archivePrefix = "arXiv",
    primaryClass = "hep-ex",
    doi = "10.1038/s41567-019-0494-8",
    journal = "Nature Phys.",
    volume = "15",
    pages = "631--634",
    year = "2019"
}

@article{Bijnens:2014lea,
    author = "Bijnens, Johan and Ecker, Gerhard",
    title = "{Mesonic low-energy constants}",
    eprint = "1405.6488",
    archivePrefix = "arXiv",
    primaryClass = "hep-ph",
    reportNumber = "LU-TP-14-16, UWTHPH-2014-10",
    doi = "10.1146/annurev-nucl-102313-025528",
    journal = "Ann. Rev. Nucl. Part. Sci.",
    volume = "64",
    pages = "149--174",
    year = "2014"
}

@article{Salone:2022lpt,
    author = "Salone, Nora and Adlarson, Patrik and Batozskaya, Varvara and Kupsc, Andrzej and Leupold, Stefan and Tandean, Jusak",
    title = "{Study of CP violation in hyperon decays at super-charm-tau factories with a polarized electron beam}",
    eprint = "2203.03035",
    archivePrefix = "arXiv",
    primaryClass = "hep-ph",
    doi = "10.1103/PhysRevD.105.116022",
    journal = "Phys. Rev. D",
    volume = "105",
    number = "11",
    pages = "116022",
    year = "2022"
}

@article{Fuchs:2003qc,
    author = "Fuchs, T. and Gegelia, J. and Japaridze, G. and Scherer, S.",
    title = "{Renormalization of relativistic baryon chiral perturbation theory and power counting}",
    eprint = "hep-ph/0302117",
    archivePrefix = "arXiv",
    reportNumber = "MKPH-T-03-2",
    doi = "10.1103/PhysRevD.68.056005",
    journal = "Phys. Rev. D",
    volume = "68",
    pages = "056005",
    year = "2003"
}

@article{Geng:2013xn,
    author = "Geng, Lisheng",
    title = "{Recent developments in SU(3) covariant baryon chiral perturbation theory}",
    eprint = "1301.6815",
    archivePrefix = "arXiv",
    primaryClass = "nucl-th",
    doi = "10.1007/s11467-013-0327-7",
    journal = "Front. Phys. (Beijing)",
    volume = "8",
    pages = "328--348",
    year = "2013"
}

@article{BESIII:2021ypr,
    author = "Ablikim, Medina and others",
    collaboration = "BESIII",
    title = "{Probing CP symmetry and weak phases with entangled double-strange baryons}",
    eprint = "2105.11155",
    archivePrefix = "arXiv",
    primaryClass = "hep-ex",
    doi = "10.1038/s41586-022-04624-1",
    journal = "Nature",
    volume = "606",
    number = "7912",
    pages = "64--69",
    year = "2022"
}

@article{Gasser:1984gg,
    author = "Gasser, J. and Leutwyler, H.",
    title = "{Chiral Perturbation Theory: Expansions in the Mass of the Strange Quark}",
    reportNumber = "CERN-TH-3798",
    doi = "10.1016/0550-3213(85)90492-4",
    journal = "Nucl. Phys. B",
    volume = "250",
    pages = "465--516",
    year = "1985"
}

@article{Jenkins:1991ts,
    author = "Jenkins, Elizabeth Ellen",
    title = "{Baryon masses in chiral perturbation theory}",
    reportNumber = "UCSD-PTH-91-12",
    doi = "10.1016/0550-3213(92)90203-N",
    journal = "Nucl. Phys. B",
    volume = "368",
    pages = "190--203",
    year = "1992"
}

@article{Yao:2016vbz,
    author = "Yao, De-Liang and Siemens, D. and Bernard, V. and Epelbaum, E. and Gasparyan, A. M. and Gegelia, J. and Krebs, H. and Mei\ss{}ner, Ulf-G.",
    title = "{Pion-nucleon scattering in covariant baryon chiral perturbation theory with explicit Delta resonances}",
    eprint = "1603.03638",
    archivePrefix = "arXiv",
    primaryClass = "hep-ph",
    doi = "10.1007/JHEP05(2016)038",
    journal = "{JHEP}",
    volume = "05",
    pages = "038",
    year = "2016"
}

@article{Alvarado:2022bok,
    author = "Alvarado, Fernando and Alvarez-Ruso, L.",
    title = "{The light-quark mass dependence of the nucleon axial charge}",
    doi = "10.31349/SuplRevMexFis.3.0308095",
    journal = "Rev. Mex. Fis. Suppl.",
    volume = "3",
    number = "3",
    pages = "0308095",
    year = "2022"
}

@article{Wang:2024jyk,
    author = "Wang, En and Geng, Li-Sheng and Wu, Jia-Jun and Xie, Ju-Jun and Zou, Bing-Song",
    title = "{Review of the Low-Lying Excited Baryons $\Sigma^*(1/2^{-})$}",
    eprint = "2406.07839",
    archivePrefix = "arXiv",
    primaryClass = "hep-ph",
    doi = "10.1088/0256-307X/41/10/101401",
    journal = "Chin. Phys. Lett.",
    volume = "41",
    number = "10",
    pages = "101401",
    year = "2024"
}

@article{Nishibuchi:2022zfo,
    author = "Nishibuchi, Takuma and Hyodo, Tetsuo",
    title = "{Nature of excited $\Xi$ baryons with threshold effects}",
    eprint = "2208.14608",
    archivePrefix = "arXiv",
    primaryClass = "hep-ph",
    doi = "10.1051/epjconf/202227110002",
    journal = "EPJ Web Conf.",
    volume = "271",
    pages = "10002",
    year = "2022"
}

@article{Banerjee:1995wz,
    author = "Banerjee, M. K. and Milana, J.",
    title = "{Decuplet reexamined in chiral perturbation theory}",
    eprint = "hep-ph/9508340",
    archivePrefix = "arXiv",
    reportNumber = "DOE-ER-40762-065, UMPP-96-19",
    doi = "10.1103/PhysRevD.54.5804",
    journal = "Phys. Rev. D",
    volume = "54",
    pages = "5804--5811",
    year = "1996"
}

@article{Borasoy:1996bx,
    author = "Borasoy, B. and Mei{\ss}ner, Ulf-G.",
    title = "{Chiral Expansion of Baryon Masses and \ensuremath{\sigma}-Terms}",
    eprint = "hep-ph/9607432",
    archivePrefix = "arXiv",
    reportNumber = "TK-96-14",
    doi = "10.1006/aphy.1996.5630",
    journal = "Annals Phys.",
    volume = "254",
    pages = "192--232",
    year = "1997"
}

@article{CBELSATAPS:2015kka,
    author = "Sokhoyan, V. and others",
    collaboration = "CBELSA/TAPS",
    title = "{High-statistics study of the reaction $\gamma p\to p\;2\pi^0$}",
    eprint = "1507.02488",
    archivePrefix = "arXiv",
    primaryClass = "nucl-ex",
    doi = "10.1140/epja/i2015-15187-7",
    journal = "Eur. Phys. J. A",
    volume = "51",
    number = "8",
    pages = "95",
    year = "2015",
    note = "[Erratum: Eur.Phys.J.A 51, 187 (2015)]"
}

@article{Sarantsev:2019xxm,
    author = "Sarantsev, A. V. and Matveev, M. and Nikonov, V. A. and Anisovich, A. V. and Thoma, U. and Klempt, E.",
    title = "{Hyperon II: Properties of excited hyperons}",
    eprint = "1907.13387",
    archivePrefix = "arXiv",
    primaryClass = "nucl-ex",
    doi = "10.1140/epja/i2019-12880-5",
    journal = "Eur. Phys. J. A",
    volume = "55",
    number = "10",
    pages = "180",
    year = "2019"
}

@article{Epelbaum:2014efa,
    author = "Epelbaum, E. and Krebs, H. and Mei\ss{}ner, U.-G.",
    title = "{Improved chiral nucleon-nucleon potential up to next-to-next-to-next-to-leading order}",
    eprint = "1412.0142",
    archivePrefix = "arXiv",
    primaryClass = "nucl-th",
    doi = "10.1140/epja/i2015-15053-8",
    journal = "Eur. Phys. J. A",
    volume = "51",
    number = "5",
    pages = "53",
    year = "2015"
}

@article{Granados:2017cib,
    author = "Granados, Carlos and Leupold, Stefan and Perotti, Elisabetta",
    title = "{The electromagnetic Sigma-to-Lambda hyperon transition form factors at low energies}",
    eprint = "1701.09130",
    archivePrefix = "arXiv",
    primaryClass = "hep-ph",
    doi = "10.1140/epja/i2017-12324-4",
    journal = "Eur. Phys. J. A",
    volume = "53",
    number = "6",
    pages = "117",
    year = "2017"
}

@article{Ando:2006xy,
    author = "Ando, Shung-ichi and Fearing, Harold W.",
    title = "{Ordinary muon capture on a proton in manifestly Lorentz invariant baryon chiral perturbation theory}",
    eprint = "hep-ph/0608195",
    archivePrefix = "arXiv",
    reportNumber = "TRI-PP-06-11",
    doi = "10.1103/PhysRevD.75.014025",
    journal = "Phys. Rev. D",
    volume = "75",
    pages = "014025",
    year = "2007"
}

@article{BESIII:2023jhj,
    author = "Ablikim, M. and others",
    collaboration = "BESIII",
    title = "{Investigation of the $\Delta I=1/2$ Rule and Test of CP Symmetry through the Measurement of Decay Asymmetry Parameters in $\Xi^-$ Decays}",
    eprint = "2309.14667",
    archivePrefix = "arXiv",
    primaryClass = "hep-ex",
    doi = "10.1103/PhysRevLett.132.101801",
    journal = "Phys. Rev. Lett.",
    volume = "132",
    number = "10",
    pages = "101801",
    year = "2024"
}

@article{BESIII:2023drj,
    author = "Ablikim, Medina and others",
    collaboration = "BESIII",
    title = "{Tests of CP symmetry in entangled $\Xi^0\overline\Xi^0$ pairs}",
    eprint = "2305.09218",
    archivePrefix = "arXiv",
    primaryClass = "hep-ex",
    doi = "10.1103/PhysRevD.108.L031106",
    journal = "Phys. Rev. D",
    volume = "108",
    number = "3",
    pages = "L031106",
    year = "2023"
}

@article{BESIII:2023sgt,
    author = "Ablikim, M. and others",
    collaboration = "BESIII",
    title = "{Test of $C\!P$ Symmetry in Hyperon to Neutron Decays}",
    eprint = "2304.14655",
    archivePrefix = "arXiv",
    primaryClass = "hep-ex",
    doi = "10.1103/PhysRevLett.131.191802",
    journal = "Phys. Rev. Lett.",
    volume = "131",
    number = "19",
    pages = "191802",
    year = "2023"
}

@misc{BESIII:2025rgd,
    author = "Ablikim, Medina and others",
    collaboration = "BESIII",
    title = "{First measurement of the absolute branching fractions of $\Sigma^+$ nonleptonic decays and test of the $\Delta I = 1/2$ rule}",
    eprint = "2512.09628",
    archivePrefix = "arXiv",
    primaryClass = "hep-ex",
    month = "12",
    year = "2025"
}

@article{Hemmert:1997ye,
    author = "Hemmert, Thomas R. and Holstein, Barry R. and Kambor, Joachim",
    title = "{Chiral Lagrangians and delta(1232) interactions: Formalism}",
    eprint = "hep-ph/9712496",
    archivePrefix = "arXiv",
    reportNumber = "TRI-PP-97-21, ZU-TH-25-97",
    doi = "10.1088/0954-3899/24/10/003",
    journal = "J. Phys. G",
    volume = "24",
    pages = "1831--1859",
    year = "1998"
}

@article{RBC:2023ynh,
    author = "Blum, Thomas and Boyle, Peter A. and Hoying, Daniel and Izubuchi, Taku and Jin, Luchang and Jung, Chulwoo and Kelly, Christopher and Lehner, Christoph and Soni, Amarjit and Tomii, Masaaki",
    collaboration = "RBC, UKQCD",
    title = "{{\ensuremath{\Delta}}I=3/2 and {\ensuremath{\Delta}}I=1/2 channels of K{\textrightarrow}{\ensuremath{\pi}}{\ensuremath{\pi}} decay at the physical point with periodic boundary conditions}",
    eprint = "2306.06781",
    archivePrefix = "arXiv",
    primaryClass = "hep-lat",
    doi = "10.1103/PhysRevD.108.094517",
    journal = "Phys. Rev. D",
    volume = "108",
    number = "9",
    pages = "094517",
    year = "2023"
}

@article{BaryonScatteringBaSc:2023zvt,
    author = "Bulava, John and others",
    collaboration = "Baryon Scattering (BaSc)",
    title = "{Two-Pole Nature of the {\ensuremath{\Lambda}}(1405) resonance from Lattice QCD}",
    eprint = "2307.10413",
    archivePrefix = "arXiv",
    primaryClass = "hep-lat",
    reportNumber = "MIT-CTP/5579",
    doi = "10.1103/PhysRevLett.132.051901",
    journal = "Phys. Rev. Lett.",
    volume = "132",
    number = "5",
    pages = "051901",
    year = "2024"
}

@article{BESIII:2025jxt,
    author = "Ablikim, Medina and others",
    collaboration = "BESIII",
    title = "{Precision CP Symmetry Test and Polarization Analysis in $\Sigma^+$ Decays}",
    eprint = "2503.17165",
    archivePrefix = "arXiv",
    primaryClass = "hep-ex",
    doi = "10.1103/ysd5-s2gn",
    journal = "Phys. Rev. Lett.",
    volume = "135",
    number = "14",
    pages = "141804",
    year = "2025"
}

@misc{BESIII:2026hgj,
    author = "Ablikim, Medina and others",
    collaboration = "BESIII",
    title = "{Precise Measurement of Matter-Antimatter Asymmetry with Entangled Hyperon Antihyperon Pairs}",
    eprint = "2602.20524",
    archivePrefix = "arXiv",
    primaryClass = "hep-ex",
    month = "2",
    year = "2026"
}

@article{BESIII:2022qax,
    author = "Ablikim, M. and others",
    collaboration = "BESIII",
    title = "{Precise Measurements of Decay Parameters and $CP$ Asymmetry with Entangled $\Lambda-\bar{\Lambda}$ Pairs}",
    eprint = "2204.11058",
    archivePrefix = "arXiv",
    primaryClass = "hep-ex",
    doi = "10.1103/PhysRevLett.129.131801",
    journal = "Phys. Rev. Lett.",
    volume = "129",
    number = "13",
    pages = "131801",
    year = "2022"
}

@misc{BESIII:2025wxe,
    author = "Ablikim, Medina and others",
    collaboration = "BESIII",
    title = "{Test of $CP$ Symmetry in the Neutral Decays of $\Lambda$ via $J/\psi \to \Lambda \bar \Lambda$}",
    eprint = "2510.24333",
    archivePrefix = "arXiv",
    primaryClass = "hep-ex",
    month = "10",
    year = "2025"
}

@article{Shtabovenko:2020gxv,
    author = "Shtabovenko, Vladyslav and Mertig, Rolf and Orellana, Frederik",
    title = "{FeynCalc 9.3: New features and improvements}",
    eprint = "2001.04407",
    archivePrefix = "arXiv",
    primaryClass = "hep-ph",
    reportNumber = "P3H-20-002, TTP19-020, TUM-EFT 130/19",
    doi = "10.1016/j.cpc.2020.107478",
    journal = "Comput. Phys. Commun.",
    volume = "256",
    pages = "107478",
    year = "2020"
}

@article{Shtabovenko:2016sxi,
    author = "Shtabovenko, Vladyslav and Mertig, Rolf and Orellana, Frederik",
    title = "{New Developments in FeynCalc 9.0}",
    eprint = "1601.01167",
    archivePrefix = "arXiv",
    primaryClass = "hep-ph",
    reportNumber = "TUM-EFT-71-15",
    doi = "10.1016/j.cpc.2016.06.008",
    journal = "Comput. Phys. Commun.",
    volume = "207",
    pages = "432--444",
    year = "2016"
}

@article{Mertig:1990an,
    author = "Mertig, R. and Bohm, M. and Denner, Ansgar",
    title = "{FEYN CALC: Computer algebraic calculation of Feynman amplitudes}",
    reportNumber = "PRINT-90-0639 (WURZBURG)",
    doi = "10.1016/0010-4655(91)90130-D",
    journal = "Comput. Phys. Commun.",
    volume = "64",
    pages = "345--359",
    year = "1991"
}

@article{Shtabovenko:2016whf,
    author = "Shtabovenko, Vladyslav",
    title = "{FeynHelpers: Connecting FeynCalc to FIRE and Package-X}",
    eprint = "1611.06793",
    archivePrefix = "arXiv",
    primaryClass = "physics.comp-ph",
    reportNumber = "TUM-EFT-75-15",
    doi = "10.1016/j.cpc.2017.04.014",
    journal = "Comput. Phys. Commun.",
    volume = "218",
    pages = "48--65",
    year = "2017"
}

@article{Patel:2015tea,
    author = "Patel, Hiren H.",
    title = "{Package-X: A Mathematica package for the analytic calculation of one-loop integrals}",
    eprint = "1503.01469",
    archivePrefix = "arXiv",
    primaryClass = "hep-ph",
    doi = "10.1016/j.cpc.2015.08.017",
    journal = "Comput. Phys. Commun.",
    volume = "197",
    pages = "276--290",
    year = "2015"
}

@article{Patel:2016fam,
    author = "Patel, Hiren H.",
    title = "{Package-X 2.0: A Mathematica package for the analytic calculation of one-loop integrals}",
    eprint = "1612.00009",
    archivePrefix = "arXiv",
    primaryClass = "hep-ph",
    doi = "10.1016/j.cpc.2017.04.015",
    journal = "Comput. Phys. Commun.",
    volume = "218",
    pages = "66--70",
    year = "2017"
}

\end{document}